\documentclass{aa}
\usepackage{upgreek}
\usepackage{graphicx}
\usepackage[varg]{txfonts}

\usepackage{natbib}

\usepackage{soul}
\newlength{\linwx}
\usepackage{xcolor}
\usepackage{hyperref}
\usepackage[flushleft]{threeparttable}
\usepackage{subcaption}
\usepackage{array}
\newcolumntype{L}{>{\centering\arraybackslash}m{6cm}}
\newcolumntype{w}{>{\centering\arraybackslash}m{3cm}}
\usepackage{amsmath}
 \usepackage{multirow}

\hypersetup{colorlinks=true,linkcolor=magenta,urlcolor=purple,citecolor = teal}

\graphicspath{ {./} }

\newcommand{\be}{\begin{equation}}
\newcommand{\ee}{\end{equation}}

\begin{document}

\title{The growth of super-Earths: the importance of a self-consistent treatment of disc structures and pebble accretion}

\author{Sofia Savvidou
\and Bertram Bitsch
}

\offprints{S. Savvidou,\\ \email{savvidou@mpia.de}}

\institute{
Max-Planck-Institut f\"ur Astronomie, K\"onigstuhl 17, 69117 Heidelberg, Germany
}

\date{Received date / Accepted date }

\abstract{The conditions in the protoplanetary disc are determinant for the various planet formation mechanisms. We present a framework which combines self-consistent disc structures with the calculations of the growth rates of planetary embryos via pebble accretion, in order to study the formation of Super-Earths. We first perform 2D hydrodynamical simulations of the inner discs, considering a grain size distribution with multiple chemical species and their corresponding size and composition dependent opacities. The resulting aspect ratios are almost constant with orbital distance, resulting in radially constant pebble isolation masses, the mass where pebble accretion stops. This supports the "peas-in-a-pod" constraint from the Kepler observations. The derived pebble sizes are used to calculate the growth rates of planetary embryos via pebble accretion. Discs with low levels of turbulence (expressed through the $\alpha$-viscosity) and/or high dust fragmentation velocities allow larger particles, hence lead to smaller pebble isolation masses, and the contrary. At the same time, small pebble sizes lead to low accretion rates. We find that there is a trade-off between the pebble isolation mass and the growth timescale with the best set of parameters being an $\alpha$-viscosity of $10^{-3}$ and a dust fragmentation velocity of 10 m/s, mainly for an initial gas surface density (at 1 AU) greater than 1000 $\rm{g/cm^2}$. A self-consistent treatment between the disc structures and the pebble sizes is thus of crucial importance for planet formation simulations. }


\authorrunning{Savvidou \& Bitsch}\maketitle

\section{Introduction}
\label{sec:Introduction}

Observational data have so far shown that planets of a few Earth masses are one of the most abundant groups of planets in the exoplanetary systems \citep[e.g.][]{Borucki+2010, Batalha+2013, Fressin+2013, Petigura+2013, Mulders+2018}.  Additionally, the super-Earth planets have been recently shown to be of similar sizes within the same planetary system \citep{Weiss+2018}, even though this was put into question by an other analysis \citep{Zhu+2018}. It is, therefore, of utmost interest to understand the mechanisms leading to their formation, especially given the fact that Super Earths are absent from the Solar System \citep{MartinLivio2015}.

The formation of planets is initiated by the coagulation \citep{WeidenscHilling1980,WeidenscHilling1984,Brauer+2008,Zsom+2011,Birnstiel+2012,Testi+2014}, condensation \citep{RosJohansen2013} of small dust particles or nucleation on icy particles \citep{Ros+2019}. There are several conditions which limit the dust growth, such as the fragmentation barrier, the bouncing barrier and the radial drift barrier \citep{Brauer+2007,Birnstiel+2010,Zsom+2010}. However, these barriers help in the rapid formation of millimeter- to centimeter-sized particles or pebbles, which is an essential contributor to planet formation.  

Planet formation can continue with the creation of larger bodies, e.g. by the streaming instability, where the dust particles concentrate into filaments with sufficient densities to gravitationally collapse into planetesimals \citep{YoudinGoodman2005,JohansenYoudin2007,YoudinJohansen2007,ChiangYoudin2010}.   In the classical core accretion scenario, the sufficiently large planetesimals serve as planetary embryos which accrete other planetesimals in order to form the planetary cores \citep[for a review]{Pollack+1996,Helled+2014}. However, this procedure has some drawbacks, among others the growth timescale, which can be even longer than the lifetime of the protoplanetary disc \citep{Pollack+1996,Rafikov2004,Levison+2010,LambrechtsJohansen2012,Fortier+2013,JohansenBitsch2019}, unless the dust density is significantly enhanced \citep{Kobayashi+2011}.

One of the proposed mechanisms to form super-Earths is the accretion of pebbles, the millimeter to centimeter sized particles, onto a preexisting planetesimal or protoplanet \citep{JohansenLacerda2010,OrmelKlahr2010,LambrechtsJohansen2012,Ormel+2017,JohansenLambrechts2017}.   The accretion of solids is not limited by the available material closely around the planetary seed, but it is aided by the drift of small solids, or pebbles \citep{WeidenscHilling1977}.  Pebble accretion thus acts on much shorter timescales compared to planetesimal accretion, especially in the outer disc regions \citep{LambrechtsJohansen2014,Bitsch+2015a,Lambrechts+2019,Izidoro+2019,JohansenBitsch2019}.

In this work, we will focus on the accretion of the small particles by an already formed planetary seed. The planet grows until it reaches the pebble isolation mass \citep{MorbidelliNesvorny2012,Lambrechts+2014,Ataiee+2018,Bitsch+2018,Surville+2020arXiv}, at which it carves a gap in the protoplanetary disc, thus trapping the pebbles in a pressure trap exterior to the planet. This critical mass depends on the disc parameters, mainly the aspect ratio and not on the available amount of solids. \citet{Wu2019} showed that protoplanetary cores could grow to the thermal mass in discs, which corresponds to the typical masses of planets in the mass-constrained sub-sample of the Kepler systems. \citet{Bitsch2019} expanded on this, showing that the pebble isolation mass could be the main driver for this observation, with a crucial dependence on the underlying protoplanetary disc structure.

The protoplanetary disc structure, thus, holds most of the key parameters and conditions which define the planet formation mechanisms. On the other hand, the grain size distribution and the subsequent opacity plays a very important role in the determination of the disc structure and the disc evolution \citep{Savvidou+2020}. A change in opacity directly affects the cooling rate of the disc and the new thermal structure leads to an updated grain size distribution. The opacity of the disc is then altered because the mass fractions of the grain sizes are different. This feedback loop operates until the disc reaches thermal equilibrium, which means that the resulting disc structure is heavily influenced by the grain size distribution and the opacity provided by the dust grains \citep{Savvidou+2020} and will then determine the evolution of planet formation. 

In this work, we obtain disc models from hydrodynamical simulations, including self-consistent grain distribution and opacities according to the grain sizes and compositions, following our previous work \citep{Savvidou+2020}. We thus have a framework present that allows a self consistent disc structure with the corresponding pebble sizes to calculate the growth rates of planetary embryos via pebble accretion. We focus on the innermost parts of a disc to study how the grain size distribution and the chemical compositions of the grains affect the formation of super-Earths. 

This paper is structured as follows. In Sect. \ref{sec:Methods} we discuss the different aspects of our model. Specifically, the opacity prescription and the chosen chemical compositions, the grain size distribution model, the disc parameters and the prescription for planet growth. In Sect. \ref{sec:Disc structure} we present the resulting disc structures from the self-consistent hydrodynamical simulations. We then discuss the pebble isolation masses of planets which could grow in the discs of our model and their growth timescales in Sect. \ref{sec:Planet formation}. We discuss the implications and limitations of our model in Sect. \ref{sec:Discussion} and summarize our findings in Sect. \ref{sec:Summary}.

\begin{figure*}
\centering
\begin{subfigure}{0.33\textwidth}
\includegraphics[width=1.1\textwidth]{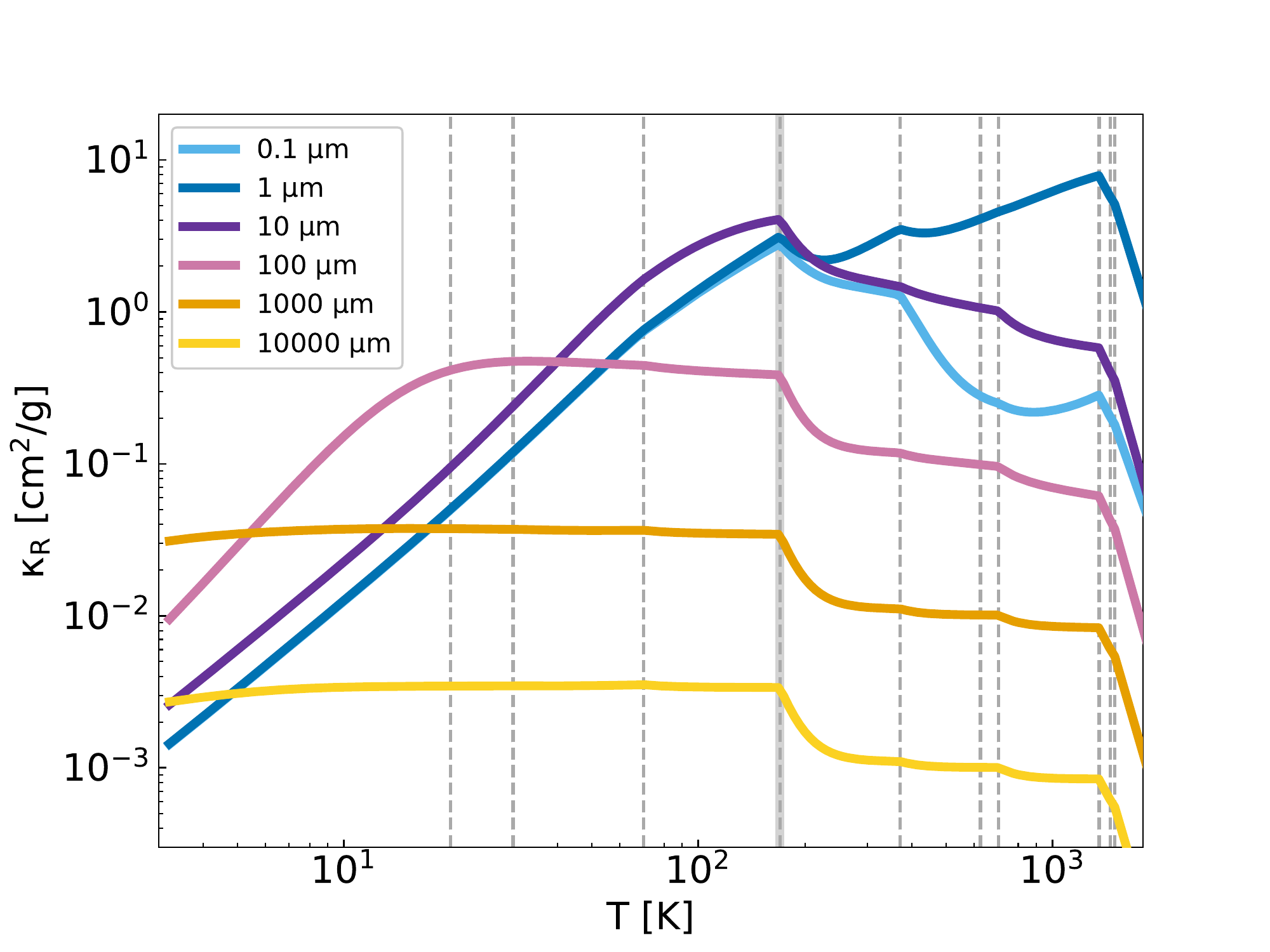}
\end{subfigure}
\begin{subfigure}{0.33\textwidth}
\includegraphics[width=1.1\textwidth]{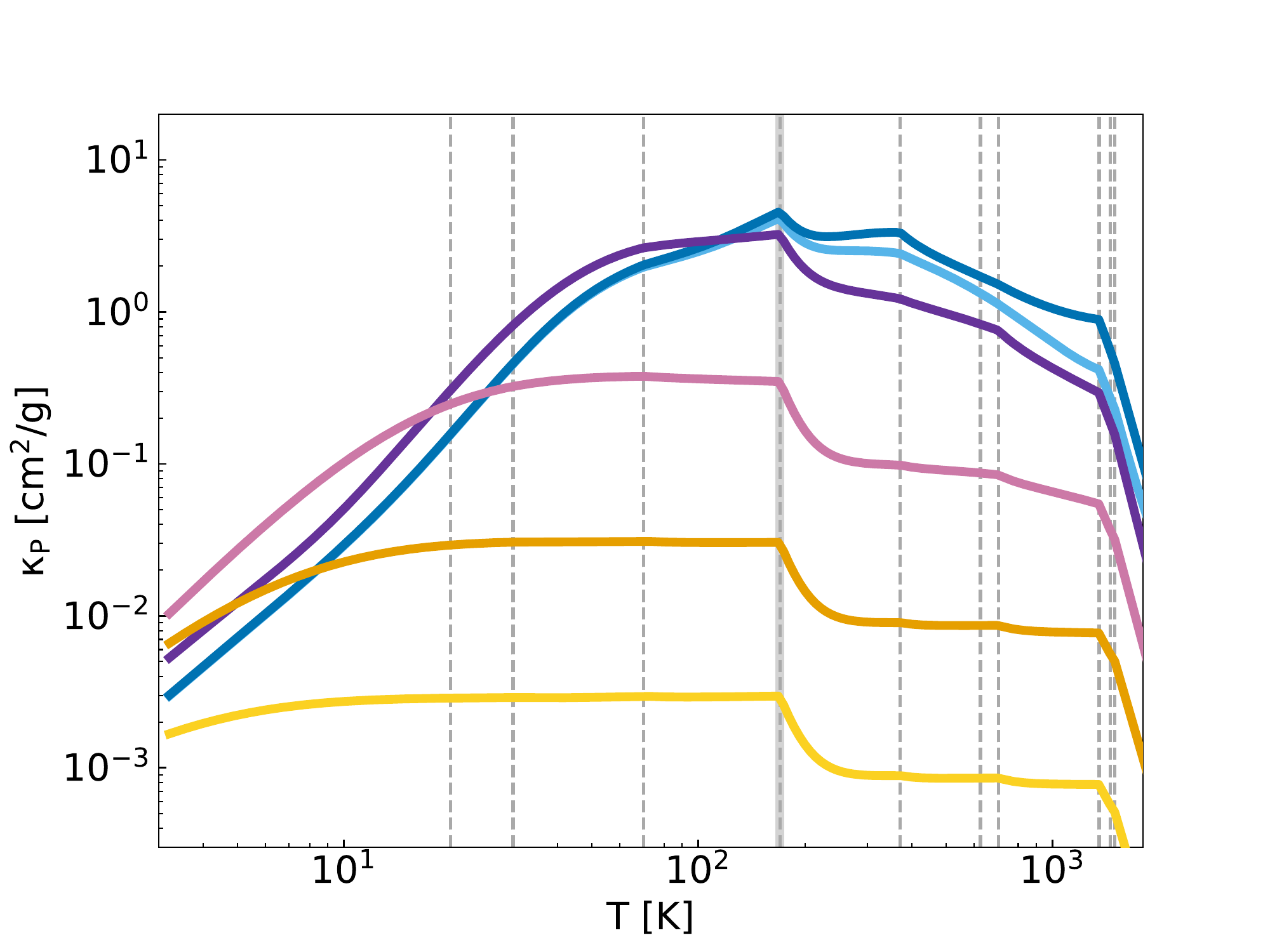}
\end{subfigure}
\begin{subfigure}{0.33\textwidth}
\includegraphics[width=1.1\textwidth]{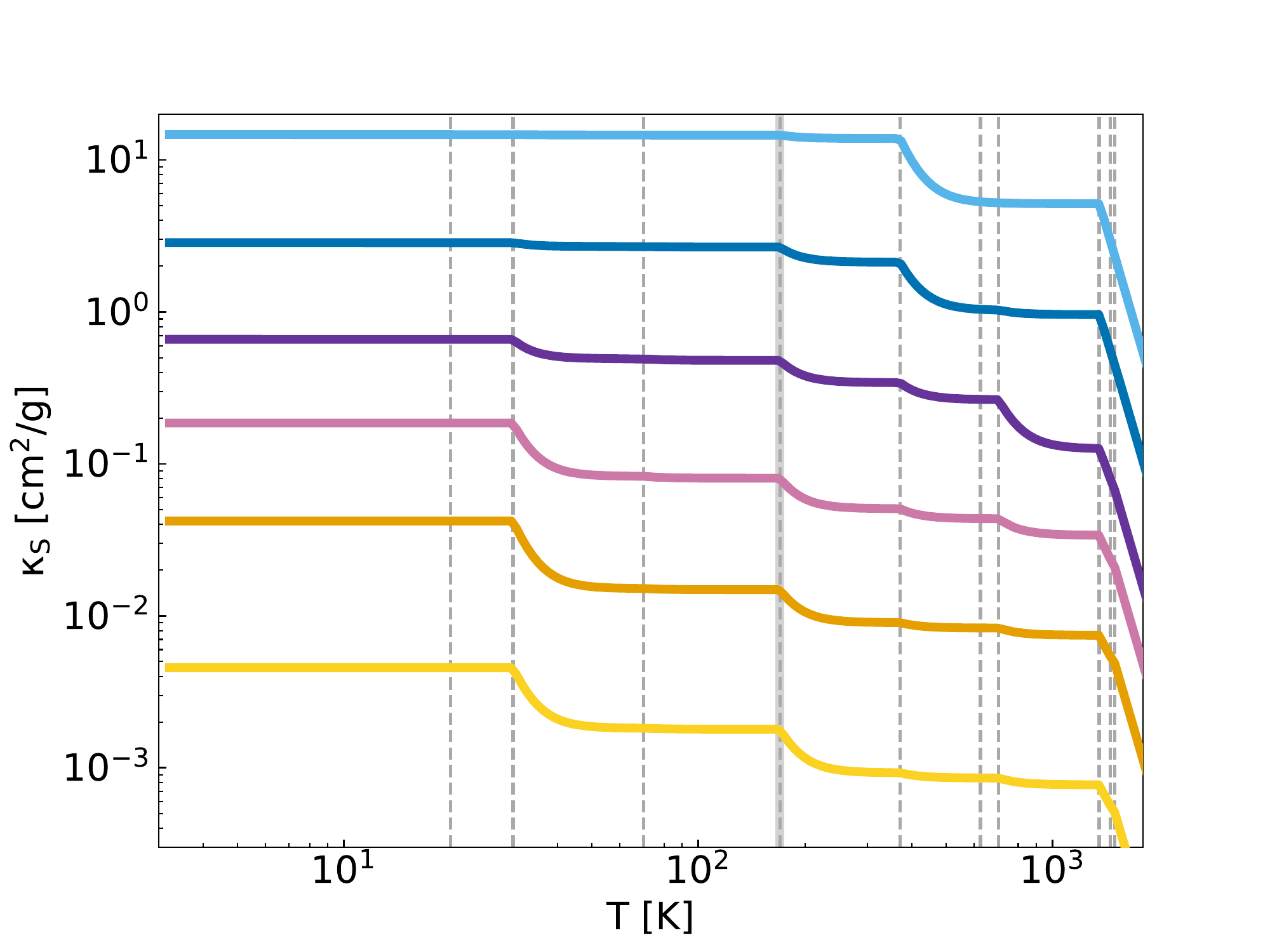}
\end{subfigure}
\caption{Rosseland, Planck, and stellar mean opacities (from left to right) as a function of temperature for grains of sizes 0.1, 1, 10, 100 $\mu$m, 1 mm, and 1 cm. The composition of the grain mixture is presented in Table \ref{tab:Grain composition}.
The gray dashed lines show the evaporation temperature of the different species, which are from the lowest to the highest temperature: $CO$, $CH_4$, $CO_2$, $H_2O$, $Fe_3O_4$, $C$, $FeS$, $Mg_2SiO_4$, $Fe_2O_3$, $MgSiO_3$. We emphasize the water-ice evaporation front with an additional gray band.}
\label{Fig:Opacities}
\end{figure*}

\section{Methods}
\label{sec:Methods}

\subsection{Opacities}
\label{subsec:Opacities}

\renewcommand{\arraystretch}{1.9}
\begin{table*}[]
\centering
\caption{Condensation temperatures and volume mixing ratios for the chemical species used in this work. We also show the densities and mass fractions considered to calculate the mean opacities as a function of temperature, along with the references of the refractive indices.}
\label{tab:Grain composition}
\begin{threeparttable}
\begin{tabular}{c|c|c|L|c|w}
\hline \hline
Species & $T_{cond}$ [K] & $\rho_s$ [$\rm{g/cm^3}$] &  Volume mixing ratios & Mass fraction & Reference for refractive indices \\ \hline
CO   & 20      &    1.0288    & 0.45 $\times$ C/H	&  21.68\% &  1,$\diamond$  \\
CH$_4$   & 30       &  0.47  &	0.25 $\times$ C/H 	& 6.882\%  &  1,$\diamond$  \\ 
CO$_2$   & 70       &  1.5  &	0.1 $\times$ C/H & 7.574\% & 1,$\diamond$  \\
H$_2$O & 170 & 1.0 & O/H - (3 $\times$ MgSiO$_3$/H + 4 $\times$ Mg$_2$SiO$_4$/H + CO/H + 2 $\times$ CO$_2$/H + 3 $\times$ Fe$_2$O$_3$/H + 4 $\times$ Fe$_3$O$_4$/H) &  21.28\%  & 2,$\bullet$           \\
Fe$_3$O$_4$    & 371      &   5.0  & (1/6) $\times$ (Fe/H - S/H)	& 4.231\% & 3,$\star$  \\
C   & 626      & 2.1 & 0.2 $\times$ C/H	&  4.129\% & 4  \\
FeS   & 704      &  4.84  & S/H	&  6.908\% &  5,$\star$ \\
Mg$_2$SiO$_4$   & 1354     &   3.275  & Mg/H - Si/H	&  13.45\% &  6,$\star$                \\
Fe$_2$O$_3$    & 1357     &   5.24   & 0.25 $\times$ (Fe/H -S/H) & 4.377\% &  3,$\star$ \\
MgSiO$_3$   & 1500     &  3.2   & Mg/H - 2 $\times$ (Mg/H - Si/H) &  9.489\%	&	6,$\star$	\\
\hline \hline                       
\end{tabular}
  \begin{tablenotes}
      \small
      \item References: [1] \citet{Hudgins+1993}, [2] \citet{Warren+2008}, [3] Amaury H.M.J. Triaud; unpublished, [4] \citet{Preibisch+1993}, [5] \citet{HenningMutschke1997}, [6] \citet{Jaeger+2003}
   \item   Data can be found in: [$\diamond$] \url{http://vizier.u-strasbg.fr}, [$\bullet$] \url{https://atmos.uw.edu/ice_optical_constants/}, [$\star$] \url{https://www.astro.uni-jena.de/Laboratory/OCDB/}
    \end{tablenotes}
\end{threeparttable}
\end{table*}

We calculate the mean Rosseland, Planck, and stellar opacities as a function of temperature using the RADMC-3D\footnote{\url{http://www.ita.uni-heidelberg.de/~dullemond/software/radmc-3d/}} code. These mean opacities are used in the energy equations of the protoplanetary disc model. In particular, the energy time evolution utilizes the Planck mean opacity which uses as a weighting function the Planck black body radiation energy density distribution ($B_{\lambda} (\lambda, T)$). The radiation flux is inversely proportional to the Rosseland mean opacity, which is calculated using the temperature derivative of the Planck distribution ($\partial B_{\lambda} (\lambda, T)/ \partial T$). Therefore, in contrast to the Planck mean opacity, the Rosseland mean opacity describes well the optically thick regions. 

For both of the aforementioned opacities the temperature taken into account is the local disc temperature. However, when we want to describe the absorption of stellar photons, we use the stellar mean opacities, which are calculated in a similar manner to the Planck mean opacities, taking into account the temperature of the stellar radiation ($B_{\lambda} (T_{\star})$), assuming isotropic radiation), which in our models is $T_{\star}$ = 4370 K. 

In contrast to our previous work \citep{Savvidou+2020}, where only water-ice and silicates were used, we have included here all of the major rock- and ice- forming species. We present them, along with their volume mixing ratios and mass fractions in Table \ref{tab:Grain composition}, following \citet{BitschBattistini2020}. We discuss in detail how we obtained these opacities in Appendix \ref{app:Refractive indices}. In Fig. \ref{Fig:Opacities} we show the opacity as a function of temperature for six representative grain sizes. The gray, vertical, dashed lines correspond to the evaporation fronts for the species we include in our model. After each evaporation front, the corresponding species sublimates and no longer contributes to the overall opacity, so the mass fractions are adjusted accordingly.

\subsection{Hydrodynamical simulations}
\label{subsec:Hydrodynamical simulations}

The viscosity in the simulations follows an $\alpha$ prescription \citep{ShakuraSunyaev1973}. We run simulations with three $\alpha$-viscosity values, namely $\alpha = 10^{-3}$, $5\times10^{-4}$ and $10^{-4}$ and list all parameters of the model in Table \ref{tab:Simulation_parameters}. 
 
 The gas surface density follows a profile 
 \be \Sigma_{g} = \Sigma_{g,0}\cdot (r/AU)^{-p}~,\ee
 with  $p = 1/2$. For each $\alpha$-viscosity we run a set of four initial gas surface densities, $\Sigma_{g,0}$ = 100, 500, 1000 and 2000 g/cm$^2$. 
 
 In all of the simulations, the stellar mass is M$_\star$ = 1 M$_\odot$, the stellar temperature is T$_\star$ = 4370 K, and the stellar radius is R$_\star$ = 1.5 R$_\odot$.  The total dust-to-gas ratio is $f_{DG} = 1.5\%$. The discs extend from 0.1 to 4 AU, except for the simulations with $\Sigma_{g,0}$ = 2000 g/cm$^2$, where the inner boundary is at 0.2 or 0.3 AU. We do not include gas opacities in our model, hence this was a necessary choice to prevent the overheating of the innermost edge and a strong shadowing that would cool down the rest of the disc. The simulations run until they reach thermal equilibrium. The disc structures from the hydrodynamical simulations are used afterwards to determine the pebble isolation masses and the pebble accretion rates, which we discuss in the following sections.

\subsection{Grain size distribution}
\label{subsec:Grain size distribution}

\renewcommand{\arraystretch}{1.3}
\begin{table}[]
\centering
\begin{tabular}{c |c| c}
    \hline
    $\alpha$ & $\Sigma_{g,0}$ [$\rm{g/cm^2}$] & $u_f$ [$\rm{m/s}$] \\
    \hline
    \multirow{1.9}*{$10^{-3}$} & 100 & \multirow{2}*{1} \\
       \multirow{2.1}*{$5\times 10^{-4}$}  & 500 &                               \\
      \multirow{2.3}*{$10^{-4}$}    & 1000 & \multirow{2}*{10}  \\
 & 2000 &                               \\
    \hline 
 $M_{\star}$ &  \multicolumn{2}{c}{ 1 $\rm{M_\odot}$} \\
  \hline
  $T_{\star}$ & \multicolumn{2}{c}{4370 K} \\
   \hline
   $R_{\star}$ & \multicolumn{2}{c}{1.5 $
   \rm{M_\odot}$} \\
   \hline
   total $f_{DG}$ & \multicolumn{2}{c}{1.5\%} \\
   \hline
\end{tabular}
 \caption{Parameters used in the simulations}
 \label{tab:Simulation_parameters}
\end{table}

\begin{figure*}
\centering
\includegraphics[scale=.45]{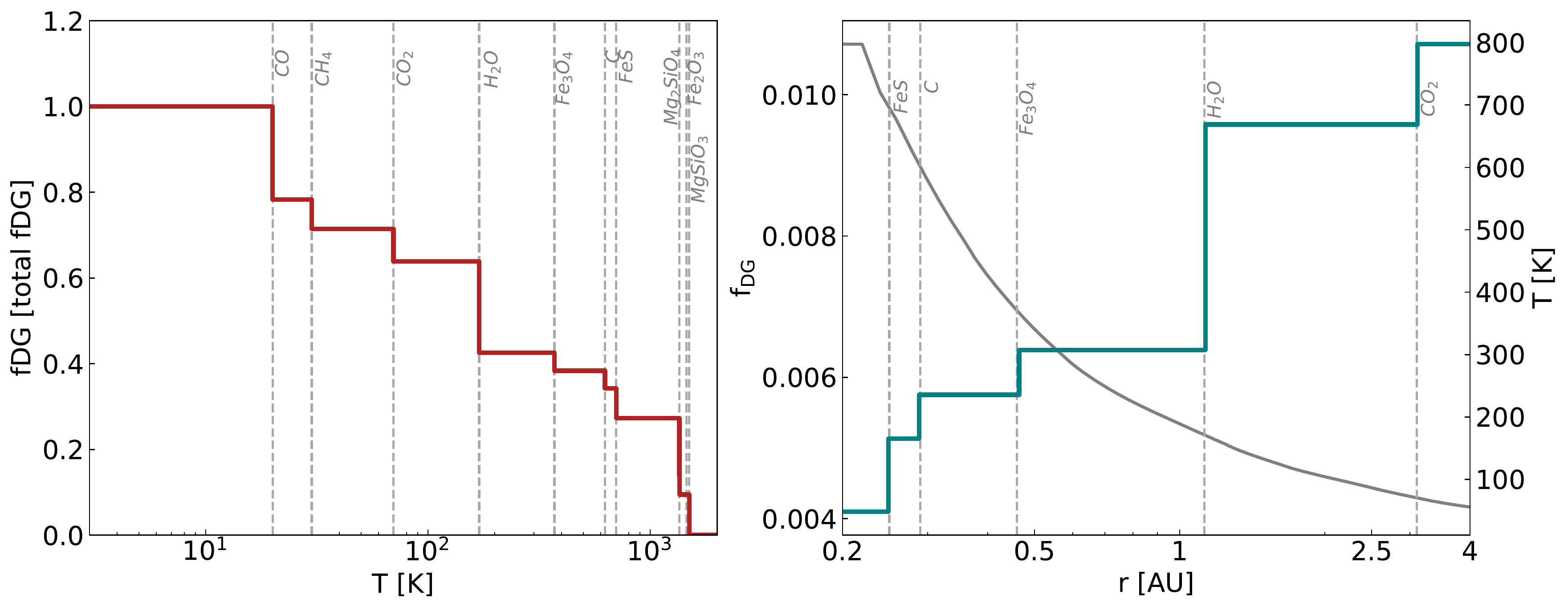}
\caption[]{Influence of the evaporation of the different chemical species on the dust-to-gas ratio in the protoplanetary disc. \emph{Left plot}: Dust-to-gas ratio in terms of the total dust-to-gas ratio in the disc as a function of temperature.  After each species evaporates the total mass fraction decreases by the corresponding mass fraction (see Table \ref{tab:Grain composition}) until all species evaporate beyond 1500 K. \emph{Right plot}: Example of the dust-to-gas ratio as a function of orbital distance for the disc with $\alpha = 10^{-3}$, $\Sigma_{g,0}=2000$ $\rm{g/cm^2}$ and $u_f=10$ m/s and a global dust-to-gas ratio of 1.5\%. Overplotted is the temperature of the disc where the evaporation fronts are easy to locate.}
\label{Fig:discfDG}
\end{figure*}

In this work we use the MRN distribution\footnote{We found that the disc structure in itself is only slightly influenced if a more complex grain size distribution \citep{Birnstiel+2011} is used \citep{Savvidou+2020}. For simplicity we use the MRN distribution.} \citep{MRN1977}, where the normalized vertically integrated dust surface densities are calculated at each orbital distance as 
\be \label{eq:MRN} \tilde{\Sigma}_{d,s_i} =\frac{s_i^{1/2}f_{DG,r}\Sigma_g}{\sum_i s_i^{1/2}},\ee
with $s_i$ the grain size and $\Sigma_g$ the gas density. In order to calculate these dust surface densities, we consider the evaporation of the chemical species as we move further-in in the disc and use $f_{DG,r}$, which is the appropriate fraction of the total dust-to-gas ratio of the disc depending on how many species have evaporated at the given location (Fig. \ref{Fig:discfDG}).

It is important to note that we do not use a constant upper boundary for the grain sizes. Instead, we use the local disc parameters to calculate the fragmentation barrier as follows
\be \label{eq:s_max} s_{max} = \frac{2\Sigma_g}{3 \pi \alpha \rho_s}\frac{u_f^2}{c_s^2}~, \ee 
with $\rho_s=1.6$ $\rm{g/cm^3}$ the assumed density of each particle, $u_f$ the fragmentation threshold velocity and $c_S$ the sound speed. 

The threshold velocity $u_f$ corresponds to the threshold after which collisions between particles always lead to fragmentation \citep{Poppe+2000}. In this work, we test two different fragmentation velocities,  $u_f = 1$ $\rm{m/s}$ and $u_f= 10$ $\rm{m/s}$, corresponding to the limits of laboratory experiments \citep{GundlachBlum2015,MusiolikWurm2019}.

\subsection{Pebble accretion}
\label{subsec:Pebble accretion}

We start the planetary seeds at 0.01 $\rm{M_{\oplus}}$, assuming they have already formed before. For simplicity we only take the Hill regime into account, which should start around this planetary mass \citep{JohansenLambrechts2017} and we follow the \citet{LambrechtsJohansen2014} accretion recipes. This implies that the accretion radius and the accretion rates of the planet are determined by the Stokes numbers of the pebbles
\be \label{eq:Stokesnumber} St_{mid} = \frac{\pi}{2}\frac{\rho_s s}{\Sigma_g}~.\ee
 We use for the planet growth only the Stokes numbers at midplane for consistency because this is the approximation we follow in our hydrodynamical models (see a more detailed discussion on the vertical distribution of the grains in \citet{Savvidou+2020}). The range of grain sizes at each location of the disc is determined by the local disc properties (see maximum grain size,  Eq. \ref{eq:s_max}).

To define the pebble accretion rate we need to distinguish between the 2D and the 3D regime, depending on how the pebble scale height $H_d$ compares to the effective accretion radius of the planet $\left(St/0.1\right)^{1/3}r_{Hill}$, where the Hill radius is 
\be r_{Hill} = r\left(\frac{M_p}{3M_{\odot}}\right)^{1/3}~. \ee
The 2D pebble accretion rate is 
\be \label{eq:Mdot_2D} \dot{M}_{\rm 2D}=2\left(\frac{St_{mid}}{0.1}\right)^{2/3}r_{Hill} v_{Hill}  \Sigma_d ~, \ee
with $v_{Hill} = \Omega r_{Hill}$, $\Omega = \sqrt{GM_{\odot}/r^3}$ and the 3D accretion rate is 
\be  \label{eq:Mdot_3D}  \dot{M}_{\rm 3D}= \dot{M}_{\rm 2D} \left[\frac{\pi\left(\frac{St_{mid}}{0.1}\right)^{1/3}r_{Hill}}{2\sqrt{2\pi}H_d}\right]~. \ee
The transition from the 2D to the 3D regime happens when 
\be \frac{\pi\left(\frac{St_{mid}}{0.1}\right)^{1/3}r_{Hill}}{2\sqrt{2\pi}H_d} < 1~, \ee
following \citet{Morbidelli+2015}. The dust scale height is derived by \citet{DubrulleSterzik1995}
 \be \label{eq:Hdust} H_d = H_g\sqrt{\frac{\alpha}{\alpha + St}}~.\ee
 We calculate the accretion rates in the 2D and 3D regime for each grain size of the distribution (Eq. \ref{eq:MRN}) and then add up all of the contribution to get the total accretion rate. 

Within this approach, we assume that the pebbles that are accreted by the planet are replenished at the planet's location by radial drift, so that the dust surface density does not change at the planets location during pebble accretion. Future work has to consider the grain drift and its influence on the disc's structure with accreting planets more accurately.                
                
\subsection{Pebble isolation mass}
\label{subsec:Pebble isolation mass}

Planetary growth halts when the planet reaches the pebble isolation mass. At this mass,  the planet has accreted enough material so that it carves a gap in the gas of the protoplanetary disc and creates a pressure bump around it \citep{PaardekooperMellema2006}. This bump prevents the dust from drifting onto the planet core and growth via pebble accretion stops
\citep{MorbidelliNesvorny2012,Lambrechts+2014,Bitsch+2018,Ataiee+2018,Surville+2020arXiv}.

The pebble isolation mass has been approximated via hydrodynamical simulations by \citet{Bitsch+2018} as 
\be \label{eq:Miso} M_{iso}=25 f_{\rm fit} M_{\oplus}~, \ee
with
\be f_{\rm fit} = \left[\frac{H/r}{0.05}\right]^3 \left[0.34\left(\frac{\log(0.001)}{\log(\alpha)}\right)^4+0.66\right]\left[1-\frac{\frac{\partial\ln P}{\partial\ln r}+2.5}{6}\right]~, \ee
where $\frac{\partial\ln P}{\partial\ln r}$ is the radial pressure gradient.
Here, we let the planets grow via pebble accretion until they reach the pebble isolation mass and track the time it takes them, to monitor if the planet can reach this mass during the disc's lifetime. However, we stop the growth at 100 Myr as this time exceeds the lifetime of protoplanetary discs quite clearly \citep[e.g.][]{Mamajek2009}.

\subsection{Advantages and limitations of our model}

Compared to 1D models, our 2D model has the advantage that the vertical structure is solved self-consistently. Furthermore, shadowing effects inside the disc caused by bumps in the disc structure that block stellar irradiation are self-consistently resolved, leading to an accurate thermal disc structure, which is not possible in 1D models. The 2D models (radial-azimuthal) of \citet{Drazkowska+2019} take into account the full coagulation and drift of particles, but are limited to the isothermal approach, thus ignoring the feedback of the grain size distribution on the thermal disc structure.

1D models, on the other hand, have the possibility to be evolved over several Myr, which is not possible with our 2D models. These 1D models \citep{Birnstiel+2012} can then accurately resolve the time evolution of the grain size distribution and additionally investigate pile-ups of material in the inner disc in time, but they can not accurately model shadowing effects. Furthermore, 1D models that self consistently takes into account grain growth, grain drift, the corresponding grain opacities and the resulting thermal structure of the disc do not exist, yet. We thus show simulations with increased dust-to-gas ratios in Sect.~\ref{sec:Planet formation} corresponding to a pile-up of grains in the inner regions caused by radial drift \citep{Brauer+2008, Birnstiel+2012}.

Our 2D approach allows us to explore the influence of the self-consistent disc models on pebble accretion (admittedly in a simplified way) and emphasize that the grain size distributions, their corresponding sizes and composition dependent opacities, affect the disc structure and with this planet formation. This effect thus needs to be included in future planet formation simulations, which, up until now, mostly operate with disc structures independently of the pebble size distribution (e.g. \citealt{Bitsch+2015b, Bitsch+2019b, JohansenLambrechts2017, Venturini+2020a}).

\section{Disc structure}
\label{sec:Disc structure}

\begin{figure*}
\centering
\includegraphics[width=\textwidth]{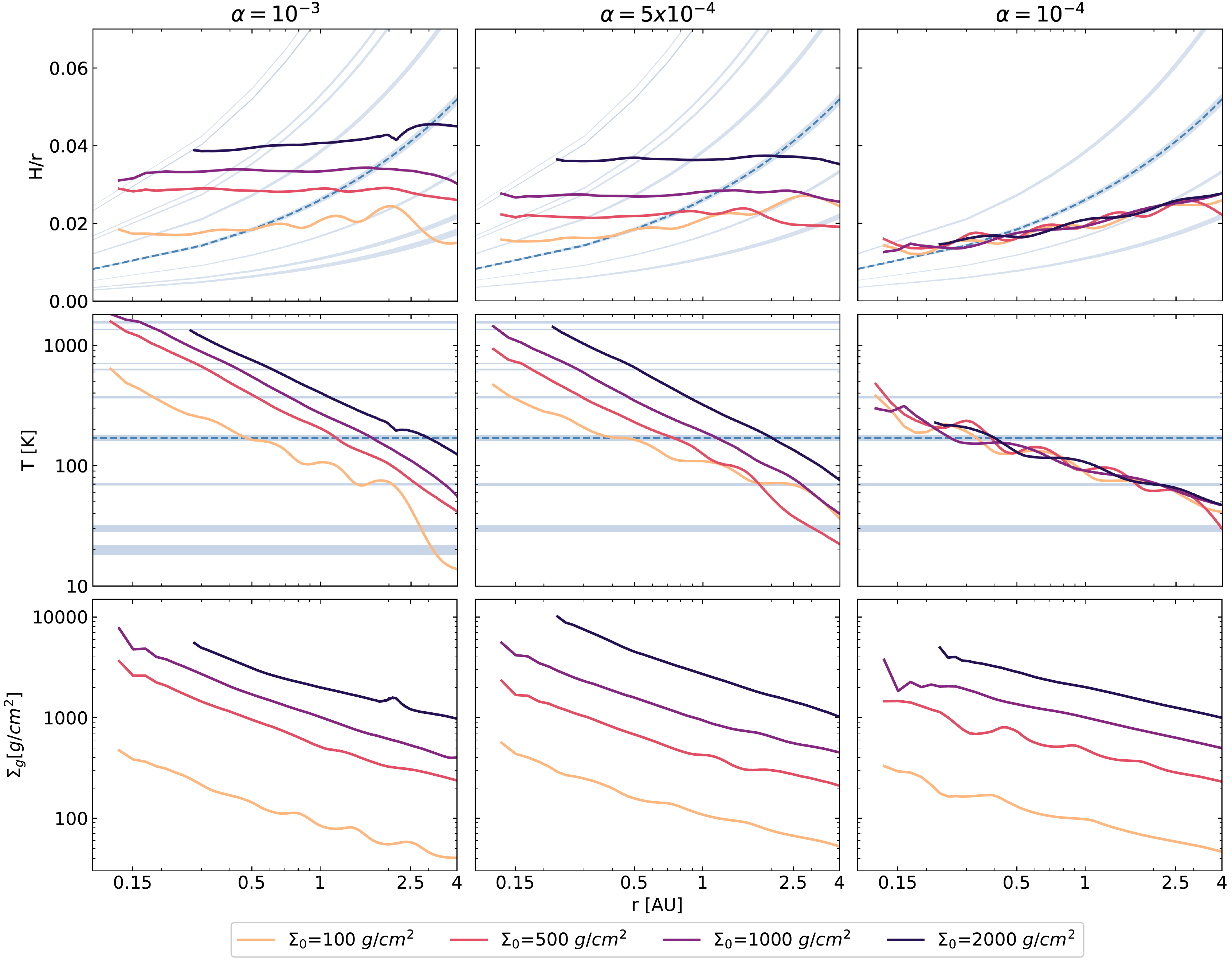}
\caption[]{Aspect ratio (upper plot), temperature (middle plot) and gas surface density (bottom plot) as a function of orbital distance for four different initial gas surface densities, from 100 g/cm$^2$ to 2000 g/cm$^2$. The turbulence parameter in viscosity is $\alpha= 10^{-3}$ in the left panel, $\alpha= 5\times10^{-4}$ in the middle panel and $\alpha= 10^{-4}$ in the right panel. The fragmentation velocity is $u_f =  1$ m/s. The light blue areas in the aspect ratio and temperature plots indicate the evaporation fronts ($\pm$ 2 K  around the evaporation temperatures of $CH_4$, $CO_2$ and $\pm$ 10 K for the other species). Starting from the outer boundary, for $\alpha =10^{-3}$ the species which evaporate are $CO$, $CH_4$, $CO_2$, $H_2O$, $Fe_3O_4$, $C$, $FeS$, $Mg_2SiO_4$, $MgSiO_3$. For  $\alpha =5\times 10^{-4}$ the evaporation fronts are of $CO_2$, $H_2O$, $Fe_3O_4$, $C$, $FeS$, $Mg_2SiO_4$, $MgSiO_3$ and for $\alpha =10^{-4}$ they are $CO_2$, $H_2O$, $Fe_3O_4$. We highlight the water ice line with an additional dashed line.}
\label{Fig:DiscStructure_uf1}
\end{figure*}

\begin{figure*}
\centering
\includegraphics[width=\textwidth]{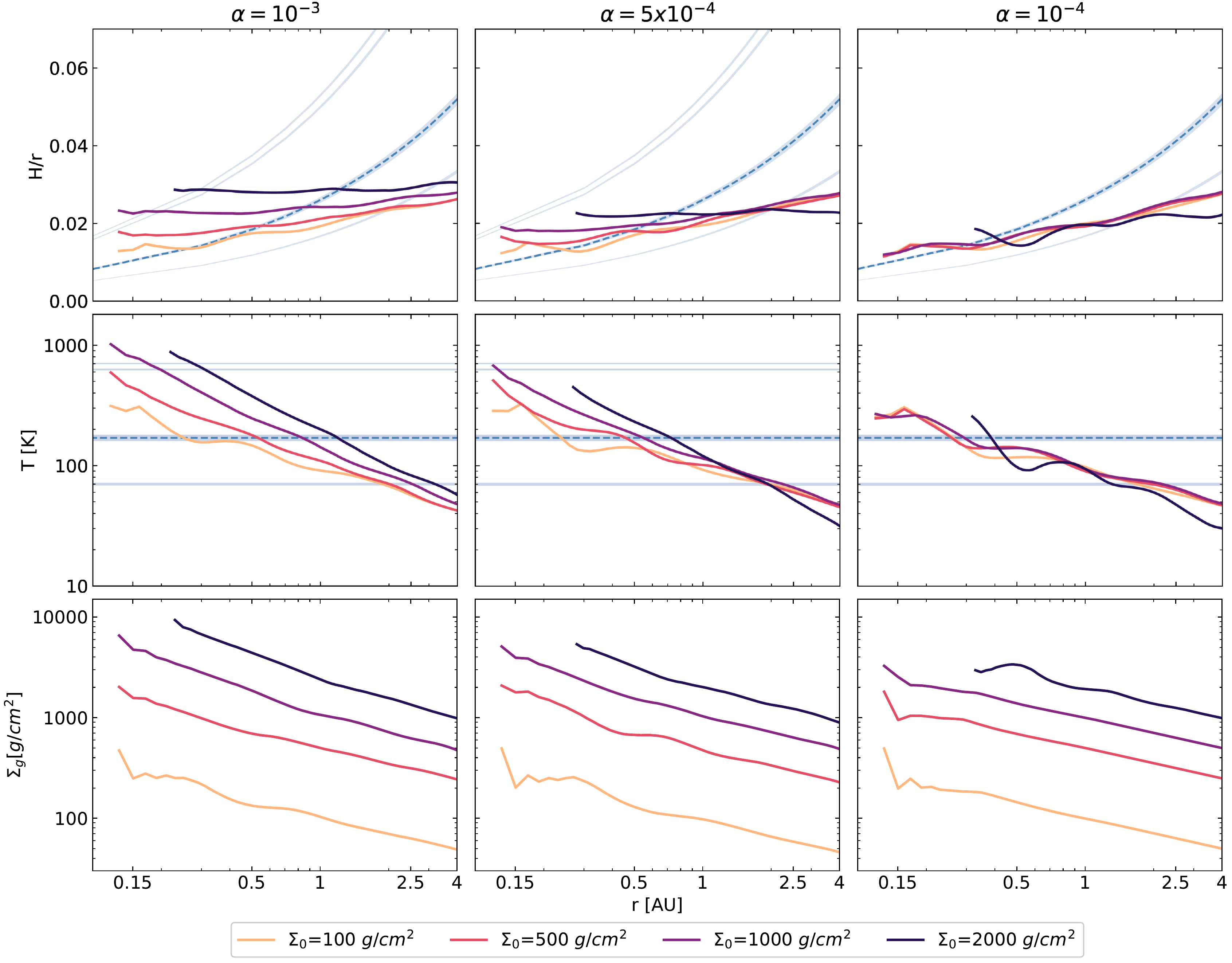}
\caption[]{Same as Fig. \ref{Fig:DiscStructure_uf1} , but using a fragmentation velocity of $u_f =  10$ m/s. Starting from the outer boundary, for $\alpha =10^{-3}$ the species which evaporate are $CO_2$, $H_2O$, $Fe_3O_4$, $C$, $FeS$. For  $\alpha =5\times 10^{-4}$ the icelines are of $CO_2$, $H_2O$, $Fe_3O_4$, $C$, $FeS$ and for $\alpha =10^{-4}$ they are $CO_2$, $H_2O$. The water icelines are also plotted with dashed lines.}
\label{Fig:DiscStructure_uf10}
\end{figure*}

\begin{figure*}
\centering
\includegraphics[width=\textwidth]{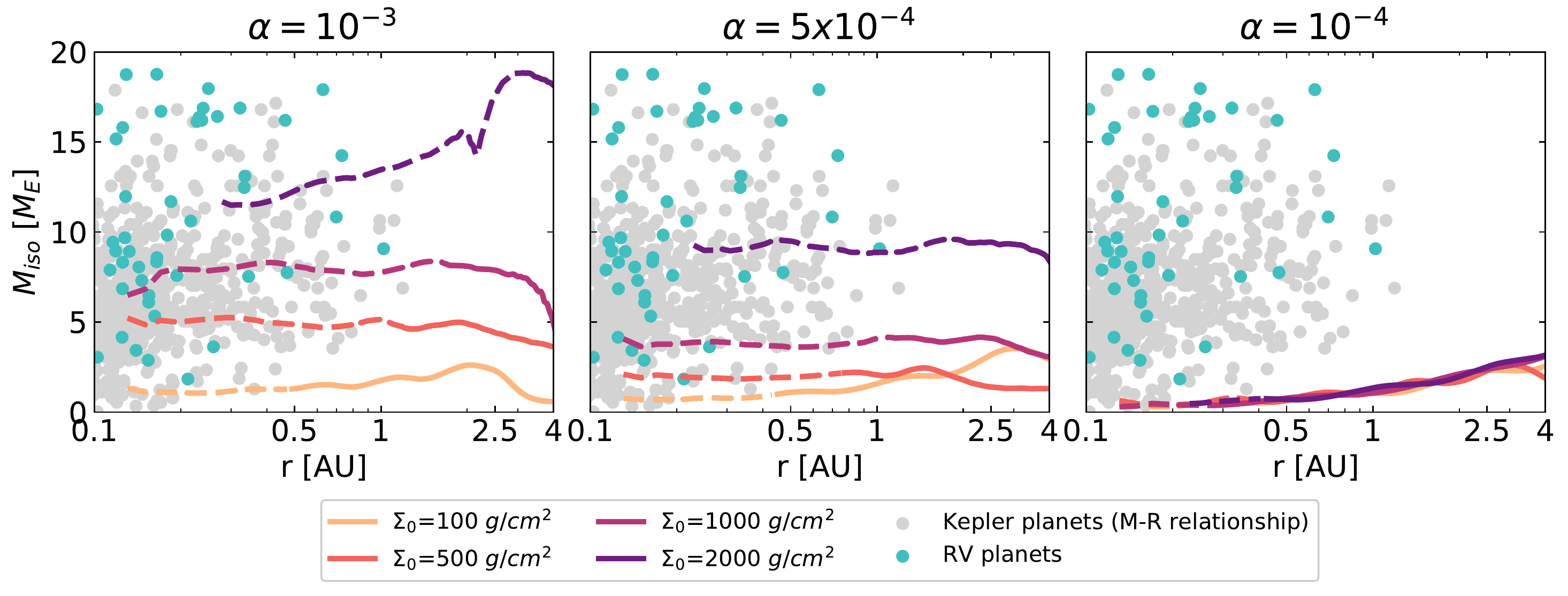}
\caption[]{Pebble isolation mass as a function of orbital distance for discs with four different initial gas surface densities, from 100 g/cm$^2$ to 2000 g/cm$^2$. The turbulence parameter in viscosity is $\alpha= 10^{-3}$ in the left panel, $\alpha= 5\times10^{-4}$ in the middle panel and $\alpha= 10^{-4}$ in the right panel. The fragmentation velocity is $u_f =  1$ m/s. We also include planetary masses of Kepler systems, with radii up to 4 $\rm{R_{\oplus}}$, derived through the \texttt{Forecaster} package \citep{ChenKipping2017} (grey circles) and super-Earths detected by RV (blue circles). The dashed lines correspond to the disc regions interior to the iceline, whereas the solid lines are the regions exterior to the iceline, where forming planets would be water rich.}
\label{Fig:Pebble_isolation_mass_uf1Z0}
\end{figure*}

\begin{figure*}
\centering
\includegraphics[width=\textwidth]{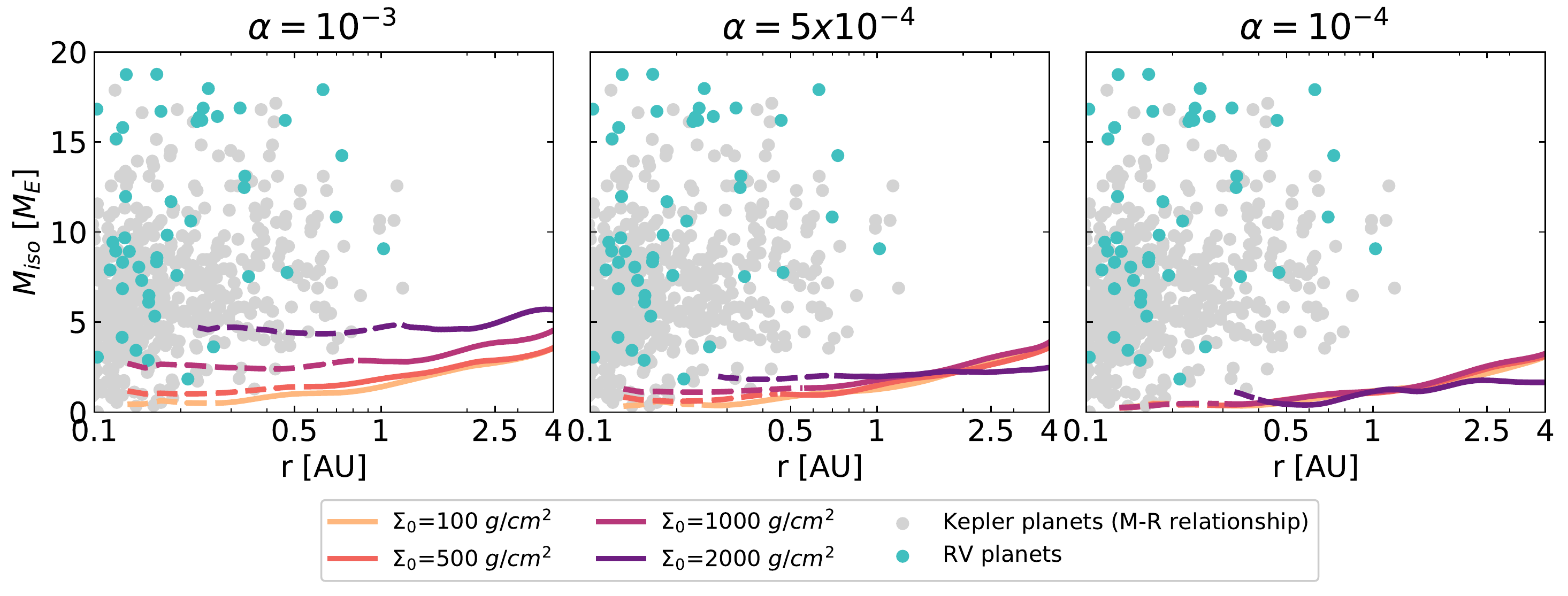}
\caption[]{Same as Fig. \ref{Fig:Pebble_isolation_mass_uf1Z0}, but for a fragmentation velocity of $u_f =  10$ m/s.}
\label{Fig:Pebble_isolation_mass_uf10Z0}
\end{figure*}

\begin{figure*}
\centering
\includegraphics[width=\textwidth]{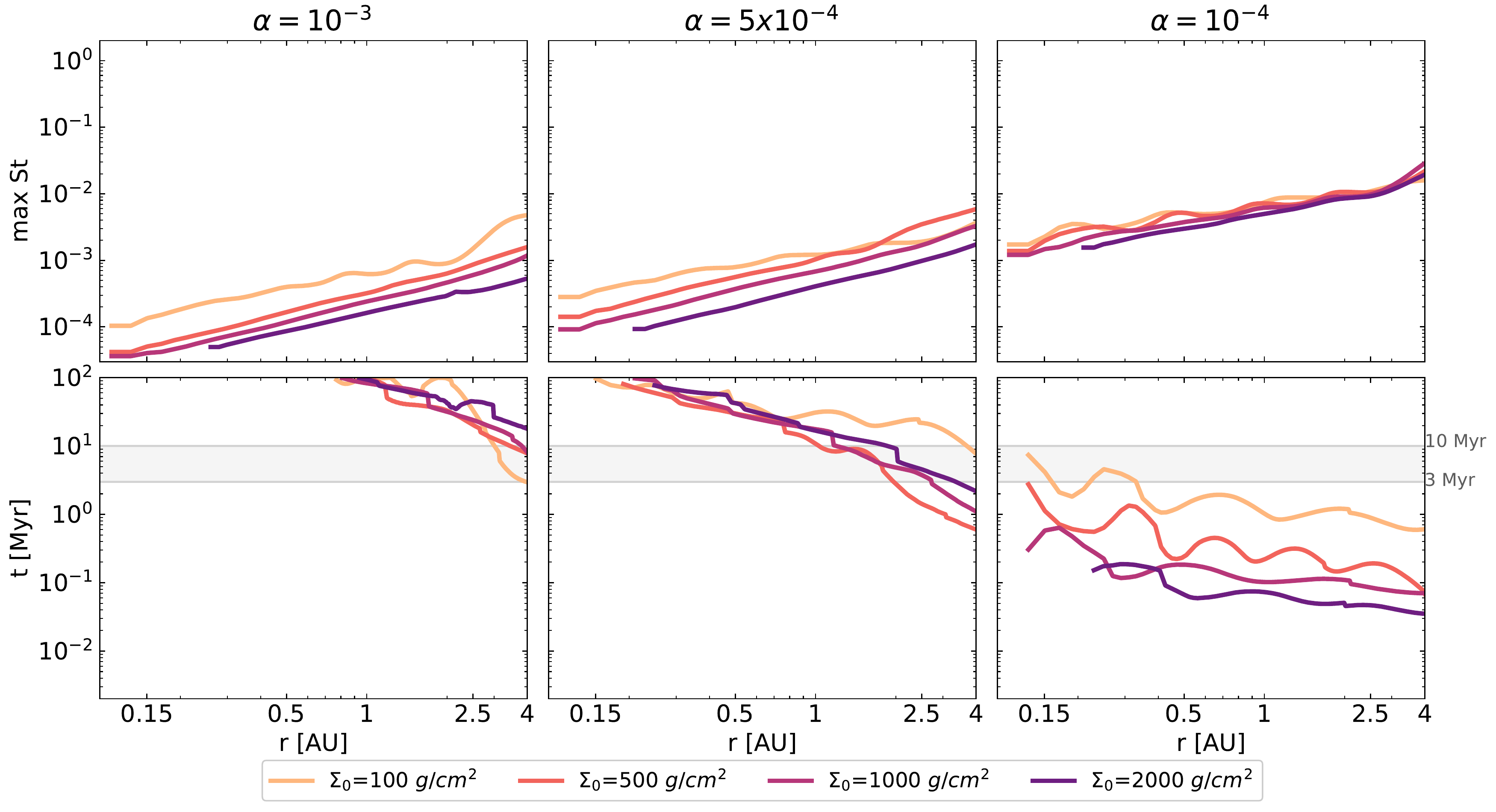}
\caption{Maximum Stokes number (top plot) and planetary growth timescale until the pebble isolation mass is reached (bottom plot) as a function of orbital distance, for the discs with the four different initial gas surface densities, $\alpha = 10^{-3}$  (left plot), $\alpha = 5\times10^{-4}$  (middle plot), $\alpha = 10^{-4}$  (right plot), and $u_f = 1$ m/s. The gray lines show the typical lifetime range of a protoplanetary disc of 3 to 10 Myr.}
\label{Fig:maxStokes&timeforMiso_uf1Z0}
\end{figure*}

\begin{figure*}
\centering
\includegraphics[width=\textwidth]{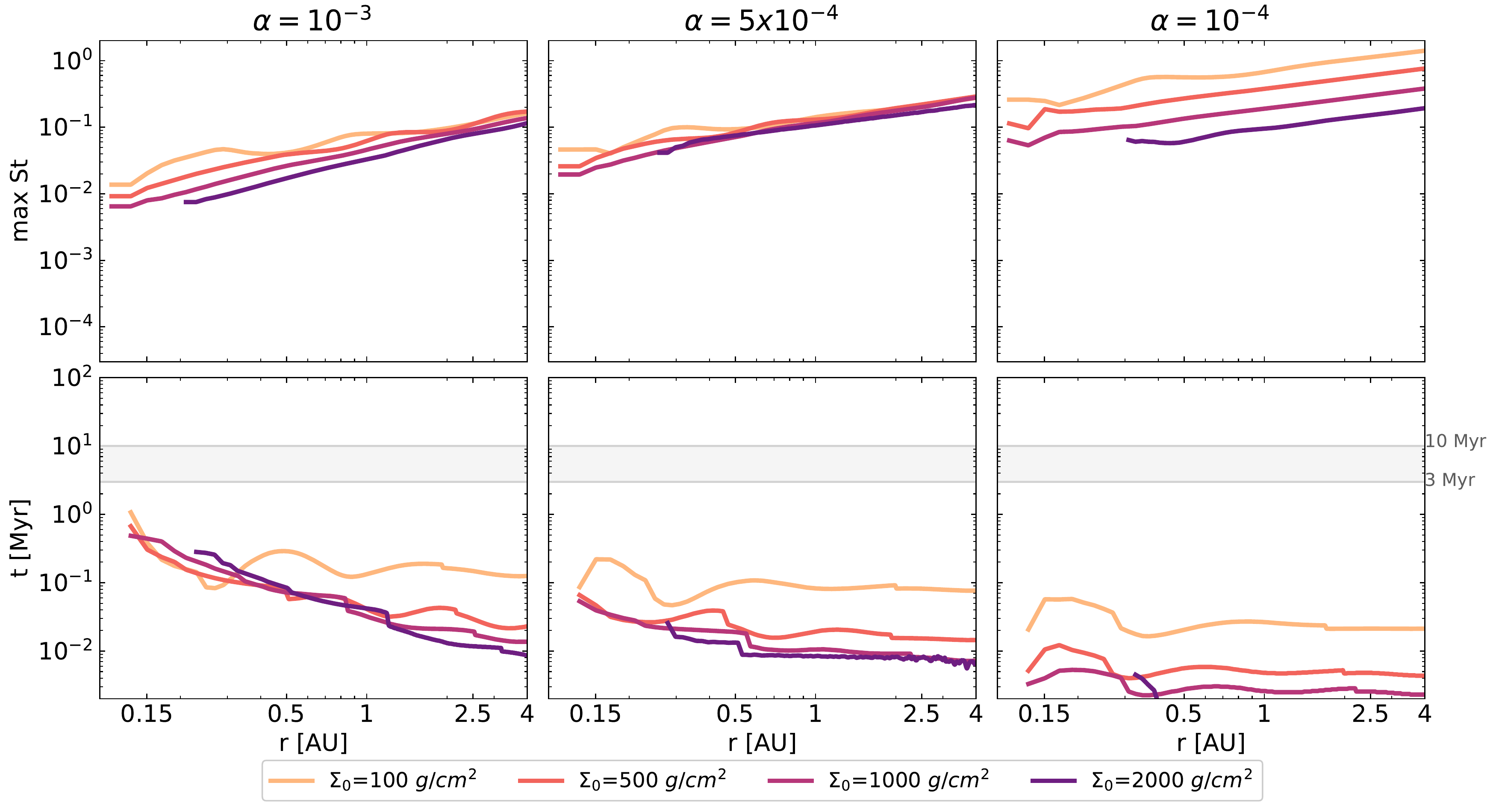}
\caption{Same as Fig. \ref{Fig:maxStokes&timeforMiso_uf1Z0} for $u_f$ = 10 m/s.}
\label{Fig:maxStokes&timeforMiso_uf10Z0}
\end{figure*}

\begin{figure}
\centering
\includegraphics[width=\columnwidth]{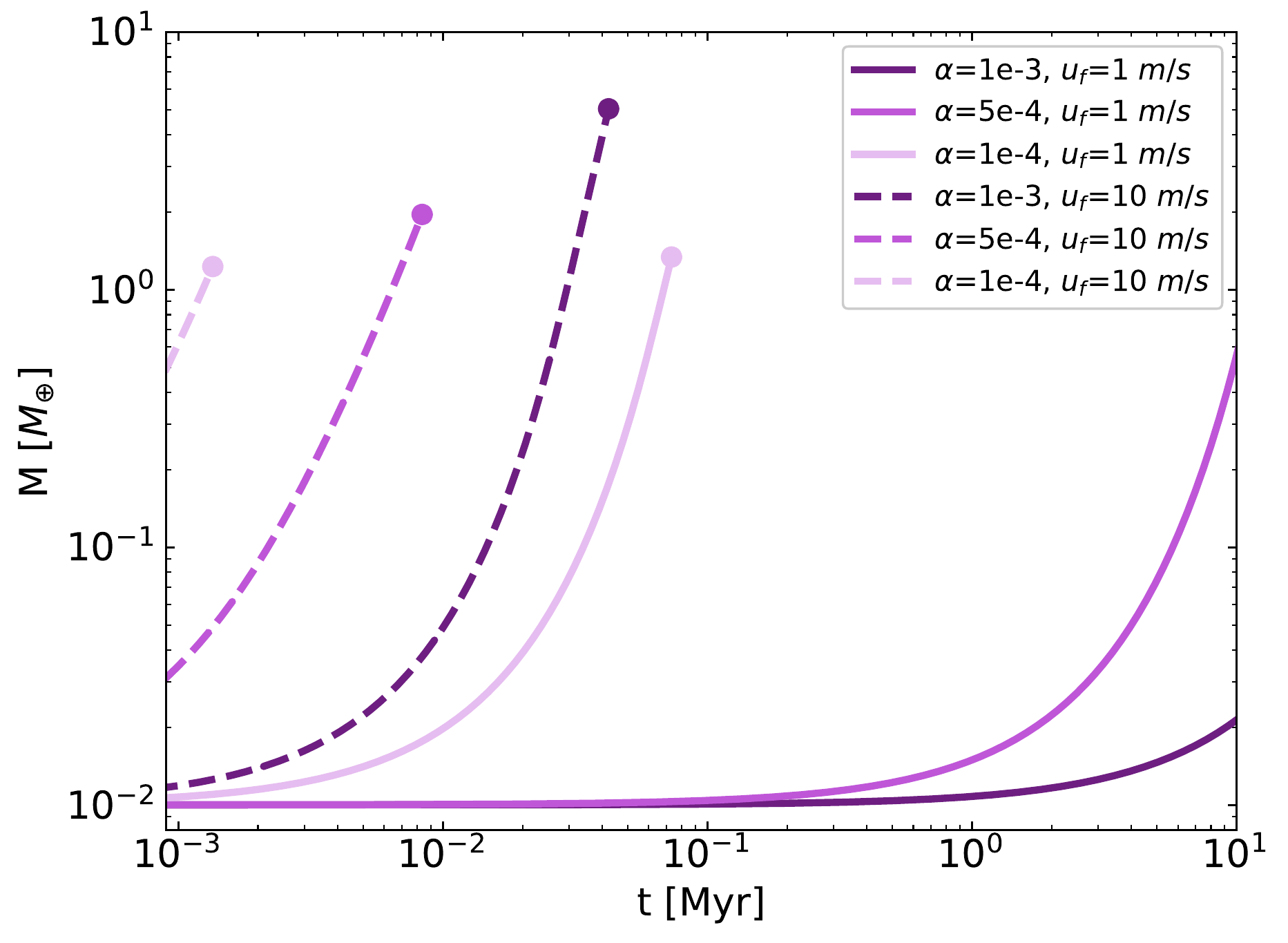}
\caption[]{Planetary mass as a function of time for the simulations with $\Sigma_{g,0}$ = 2000 g/cm$^2$, all $\alpha$-viscosities and both of the fragmentation velocities at 1 AU. The dot marks the final planetary mass determined by the pebble isolation mass (Figs.\ref{Fig:Pebble_isolation_mass_uf1Z0} and \ref{Fig:Pebble_isolation_mass_uf10Z0}). The growth time scale is longer than 10 Myr in discs with $\alpha > 5\times10^{-4}$ when particle collisions are limited by a fragmentation velocity of 1 m/s.}
\label{Fig:Pmass_Sg02000}
\end{figure}

\subsection{Dependence on viscosity, gas surface density and fragmentation velocity}

We investigate the influence of different $\alpha$-viscosities, initial gas surface density and fragmentation velocities on the grain sizes and disc structures. The specific values used are shown in Table \ref{tab:Simulation_parameters}. The influence of the different $\alpha$-viscosity and surface densities $\Sigma_{g,0}$ to the resulting disc structure has been discussed in detail in \citet{Savvidou+2020}. In a nutshell, as the $\alpha$-viscosity decreases, the disc gets colder for two reasons. Since the viscous heating decreases, the disc cools down more efficiently. At the same time, the dust particles face less destructive collisions and grow to larger sizes, which have lower opacities. This aids further the cooling of the disc. 

We present the results of our simulations with different $\alpha$ parameter and for different initial gas surface densities, utilizing a fragmentation velocity of 1 m/s in Fig. \ref{Fig:DiscStructure_uf1}. Clear trends with viscosity and gas surface density are visible.

Decreasing the gas surface density scales down the total dust surface density. As a consequence, the disc is colder, because the viscous heating is less efficient and radiative cooling is enhanced, given that the cooling is inversely proportional to the disc's density. However, for very low $\alpha$-viscosity values the difference between discs with different surface density diminishes \citep{Savvidou+2020}. We find here as well that for $\alpha = 10^{-4}$ (third column in Fig. \ref{Fig:DiscStructure_uf1}), the aspect ratio of the discs is very low and almost independent of the gas surface density. In this case, the disc is mainly optically thin and the dominant heating mechanism is stellar irradiation, resulting in very similar disc structures.

The same applies to the simulations using higher fragmentation velocities. In this case, larger grains are available and dominate the opacity of the disc, hence the cooling of the disc is very efficient (Fig. \ref{Fig:DiscStructure_uf10} for $\alpha = 5\times10^{-4}~\text{and}~10^{-4}$). Even for the highest $\alpha=10^{-3}$ (first column in Fig. \ref{Fig:DiscStructure_uf10}), the difference in the aspect ratios of discs with varying surface densities is not as pronounced as in the corresponding discs with lower fragmentation velocity (Fig. \ref{Fig:DiscStructure_uf1}). This shows that the larger grain sizes provide such low opacities that even an alpha-viscosity of $10^{-3}$ cannot sufficiently heat up the disc. As a result, the water iceline is already close to 1 AU, even for the highest gas surface density.

In Figs. \ref{Fig:DiscStructure_uf1}-\ref{Fig:DiscStructure_uf10} we see some radial variations in some of the disc profiles, especially for very low surface densities. This is caused mainly by convection, created by the vertical temperature gradient which depends on the opacity gradient in the vertical direction and is present in the optically thick regions of the disc \citep{Bitsch+2013b}. However, in this work, we have included multiple chemical species in our opacity prescription, and this creates dips in the opacity as a function of temperature at the evaporation fronts of the various species\footnote{In contrast, our previous simulations \citep{Savvidou+2020} only included 2 species, resulting in less variations in the disc profile.} (see Fig. \ref{Fig:Opacities}). Even if these dips are minimal, they can create small bumps in the aspect ratio profiles which are amplified by the convection and lead to an enhancement of these perturbations. 

The protoplanetary discs in our models, or regions within them, which are not strongly affected by convection, do not show a significant influence from the multiple evaporation fronts of the chemical species we have included, because the changes in the opacity are small. The strongest effect remains with the evaporation of water-ice, which causes a strong dip in the opacity and hence a more enhanced bump in the aspect ratio (Fig.\ref{Fig:DiscStructure_uf1} and \ref{Fig:DiscStructure_uf10}). Nevertheless, the overall opacity is slightly different compare to a prescription where only water-ice and silicates is used \citep{Savvidou+2020}. In addition to this, including multiple chemical species is important because they could be used in future work to predict the possible compositions of the planets growing within the discs of our models.

In our previous work \citep{Savvidou+2020}, we discuss the dependency of the water iceline position on the $\alpha$-viscosity, the gas surface density and the dust-to-gas ratio. We also find here that it moves further in if the gas surface density or the $\alpha$-viscosity decreases. Here, we also use $u_f = 10$ m/s and find that the higher fragmentation velocity moves the iceline inwards compared to $u_f = 1$ m/s. This has consequences, not only for the planet formation mechanisms, but also for the potential compositions of the created planets \citep{Bitsch+2019b,Schoonenberg+2019,Venturini+2020a}. A more detailed discussion about planet growth within the discs of our models follows in Sect. \ref{sec:Planet formation}.

\subsection{Disc evolution}

Our 2D hydrodynamical models can not be integrated over several Myr due to computational limitations. Instead, the disc evolution can be mimiced by reducing the overall gas surface density, where lower disc surface densities correspond to lower stellar accretion rates and thus to older discs \citep{Hartmann+1998}. As a consequence the viscous heating is reduced and the discs become colder as they age. However, assigning an absolute age to our here presented disc structures is difficult, because the exact time evolution of a disc depends on more than just the viscosity and the gas surface density, because also disc winds can drive the evolution of the disc \citep{Bai+2016, Suzuki+2016}. 

In our model, the evaporation fronts of different solids depend on the temperature (table \ref{tab:Grain composition}) and consequently are also closer to the central star in discs with lower gas surface density. At the evaporation fronts, solids evaporate and thus the solid surface density is reduced (Fig.~\ref{Fig:discfDG}), leading to lower opacities and thus more efficient cooling, which can cause bumps in the disc structure if the solid abundances change by a large factor, as e.g. at the water ice line. However, the maximal pebble size by fragmentation does not change, because the maximal pebble size does not depend on dust surface density (Eq.~\ref{eq:s_max}). As a consequence, the pebble sizes are smooth across the evaporation fronts, implying that even if disc evolution over several Myr was taken into account, it would happen smoothly and no bumps/dips will be generated at the evaporation fronts. As the disc becomes older and its surface density reduces in time, the maximal pebble size increases. We discuss the consequences of this on planet formation via pebble accretion in the next section.

\section{Planet formation}
\label{sec:Planet formation}

An important implication which can be derived from the disc structures of the presented models is that the aspect ratio profiles, independently of the parameters used, are almost constant with orbital distance. This is expected in these inner regions where viscous heating dominates over stellar irradiation which would flare up the disc \citep{ChiangGoldreich1997,DullemondDominik2004,Bitsch+2015a,Ida+2016,Savvidou+2020}.

The implication of this observation is that the planetary systems which could potentially form in those discs would have very similar masses, since the pebble isolation mass depends on the aspect ratio (see Eq. \ref{eq:Miso}). It has been recently observed among the \emph{Kepler} systems that planets within the same system have similar sizes \citep{Weiss+2018}. \citet{Millholland+2017} suggested that the "peas-in-a-pod" trend is also true for the planetary masses within the same system. 

We now compare the pebble isolation masses derived from our disc simulations with the super-Earth population and then discuss planetary growth within these disc environments. We stress here again that discs with large gas surface densities could correspond to young discs, while the discs with low gas surface densities could correspond to older discs. As the exact time evolution of discs is complicated (e.g. \citealt{Bai+2016, Suzuki+2016}) and not self-consistently included in our model, we do not link our disc structures to a time evolution, but just discuss the implications of the fixed disc structures on planetary growth via pebble accretion under the simple assumption that the disc structure does not evolve in time.

\subsection{Pebble isolation mass}
\label{subsec:Isolation mass}

We use the equilibrium disc structures from the 2D hydrodynamical simulations to study the growth of super-Earths via pebble accretion. It is assumed that planets will grow up to the pebble isolation mass because then they carve a gap in the protoplanetary disc and create a pressure bump exterior to their orbits which traps pebbles and prevents them from being accreted. We thus calculate the pebble isolation mass for each one of our disc models to find the maximum mass our planets could reach. The pebble isolation mass (Eq. \ref{eq:Miso}) depends on the disc structure and specifically, its aspect ratio, $\alpha$-viscosity and the radial pressure gradient. We present the pebble isolation mass derived from our hydrodynamical simulations in Fig. \ref{Fig:Pebble_isolation_mass_uf1Z0} and Fig. \ref{Fig:Pebble_isolation_mass_uf10Z0}, where we also overplot the masses of close-in super-Earths inferred, using a mass-radius relationship \citep{ChenKipping2017}, from the \emph{Kepler} observations as well as for planets with RV mass determinations.

The aspect ratio profiles for the disc range in our models are almost constant with orbital distance, so we expect and find a low dependency of the pebble isolation mass on the orbital distance. For $\alpha = 10^{-3}$ and $u_f$ = 1 m/s the pebble isolation mass reaches almost 19 $M_{\oplus}$ for the highest initial gas surface density, $\Sigma_{g,0} = 2000$ g/cm$^2$. With the same fragmentation velocity and $\alpha = 5\times10^{-4}$ we still get high enough isolation masses, to match the majority of the observed super-Earths / mini-Neptunes, mainly for the highest surface density, $\Sigma_{g,0}=2000$ g/cm$^2$. However, for lower $\alpha$ values or higher fragmentation velocities the pebble isolation mass hardly exceeds 3-4 $\rm{M_{\oplus}}$. This also means that with these sets of parameters it is hard to explain the bulk of the masses of close-in super-Earths / mini-Neptunes by pure pebble accretion.
  
Increasing the fragmentation velocity to $u_f = 10$ m/s, we find that the pebble isolation masses are significantly reduced. This happens because of the larger particles (Eq. \ref{eq:s_max}), which lead to a smaller aspect ratio (see Sect. \ref{sec:Disc structure} and Eq. \ref{eq:Miso} ). The highest mass we find is around 5 $\rm{M_\oplus}$ for $\alpha = 10^{-3}$ and again the highest initial gas surface density, $\Sigma_{g,0} = 2000$ g/cm$^2$. For the rest of the simulations, the pebble isolation mass is so low that the masses of the inner super-Earths might not be reached without a significant amount of collisions between the bodies.

Considering only the pebble isolation masses of the discs discussed here, we can conclude that in order to explain the inferred masses from \emph{Kepler} detections, we would need a relatively high viscosity of $\alpha = 10^{-3}$. However, it is also important to consider whether pebble accretion can operate efficiently enough so that the planets reach these masses in a timely manner. We discuss this in the following section. 

\subsection{Planet growth until pebble isolation mass}
\label{subsec:Planet_growth}

To study planet growth we calculate the pebble accretion rate using  Eqs. \ref{eq:Mdot_2D} and \ref{eq:Mdot_3D}. The Stokes numbers are determined by the MRN distribution. For simplicity we ignore planetary migration. We show the maximum Stokes numbers as a function of orbital distance in the upper plots of Fig. \ref{Fig:maxStokes&timeforMiso_uf1Z0} and \ref{Fig:maxStokes&timeforMiso_uf10Z0}  for all of the different parameters used. The Stokes numbers are inversely proportional to the gas density of the disc (Eq. \ref{eq:Stokesnumber}). The maximum Stokes numbers are hence an increasing function of the orbital distance. They also show some radial variations for the same reason.

We show the growth timescales for our different simulations in the bottom plots of Figs. \ref{Fig:maxStokes&timeforMiso_uf1Z0} and \ref{Fig:maxStokes&timeforMiso_uf10Z0}. These timescales represent the time it takes for the planets to reach the pebble isolation mass at the given orbital distance (Figs. \ref{Fig:Pebble_isolation_mass_uf1Z0} and \ref{Fig:Pebble_isolation_mass_uf10Z0}) accreting pebbles with a size distribution corresponding to the planets location.
 The growth timescales are not entirely smooth. This is related, firstly, to the variations in the Stokes number (upper plots of Figs. \ref{Fig:maxStokes&timeforMiso_uf1Z0} and \ref{Fig:maxStokes&timeforMiso_uf10Z0} for the maximum values). Secondly, there are also some variations in the dust surface density, because of the evaporation fronts (see the steps for the dust-to-gas ratio in Fig. \ref{Fig:discfDG}).

When considering the time it takes to reach the isolation masses it is important to compare it with the time of the gas dispersal, marked by a gray band in the bottom plots of Fig. \ref{Fig:maxStokes&timeforMiso_uf1Z0} and Fig. \ref{Fig:maxStokes&timeforMiso_uf10Z0}. After this event, planets can continue to grow via mutual collisions, however the formation pathway is different and if the planets have not reached sufficient masses already, they will end up being terrestrial rather that super-Earths \citep{Lambrechts+2019}. 

For fragmentation velocities of 1 m/s and $\alpha=10^{-3}$, pebble accretion is inefficient in the inner disc regions (with a slight dependence on the gas surface density, Fig. \ref{Fig:maxStokes&timeforMiso_uf1Z0}) and thus longer than the typical lifetimes of protoplanetary discs. These timescales get significantly shorter for a decrease in the $\alpha$-viscosity or an increase the fragmentation velocity to 10 m/s (Fig. \ref{Fig:maxStokes&timeforMiso_uf10Z0}) which allow larger grains to be accreted (Eq. \ref{eq:s_max}), enhancing pebble accretion (Eqs. \ref{eq:Mdot_2D} and \ref{eq:Mdot_3D}). Our simulations thus indicate that either a low $\alpha$-viscosity or larger grain fragmentation velocities are needed to allow fast enough growth via pebble accretion.

In Fig. \ref{Fig:Pmass_Sg02000} we show the planetary growth of embryos located at 1 AU as a function of time, for planets growing in discs with $\Sigma_{g,0} = 2000$ g/cm$^2$ for all $\alpha$-viscosities and fragmentation velocities. The planet grows fastest in environments with low viscosities and large fragmentation velocities. However, the pebble isolation mass is small in these cases. Only for cases of high $\alpha$ and low fragmentation velocity, is the pebble isolation mass large enough to match the observed exoplanet population.

The larger grains carry lower opacities which enhances the cooling of the disc. The low temperature also translates to a low aspect ratio (Fig. \ref{Fig:DiscStructure_uf10}), hence the isolation masses are very low (Fig. \ref{Fig:Pebble_isolation_mass_uf10Z0}) and even though the timescales to reach them are very short, the planets that could potentially form cannot explain the majority of the masses of super-Earths and mini-Neptunes. The best options for sufficiently large isolation masses and short growth timescales are  $\alpha = 10^{-3}$, $u_f=10$ m/s, which leads to masses from 0.4 to 5.7 $\rm{M_{\oplus}}$, depending on the gas surface density and the orbital distance.

In the following section we will explore if planet formation at earlier stages (high gas surface densities) and higher dust-to-gas ratios could increase the pebble isolation mass, and keep the planetary growth timescales short at the same time.

\subsection{Testing higher dust-to-gas ratio, gas surface density and initial planetary seed mass}
\label{subsec:Improvements}

\begin{figure}
\centering
\includegraphics[width=\columnwidth]{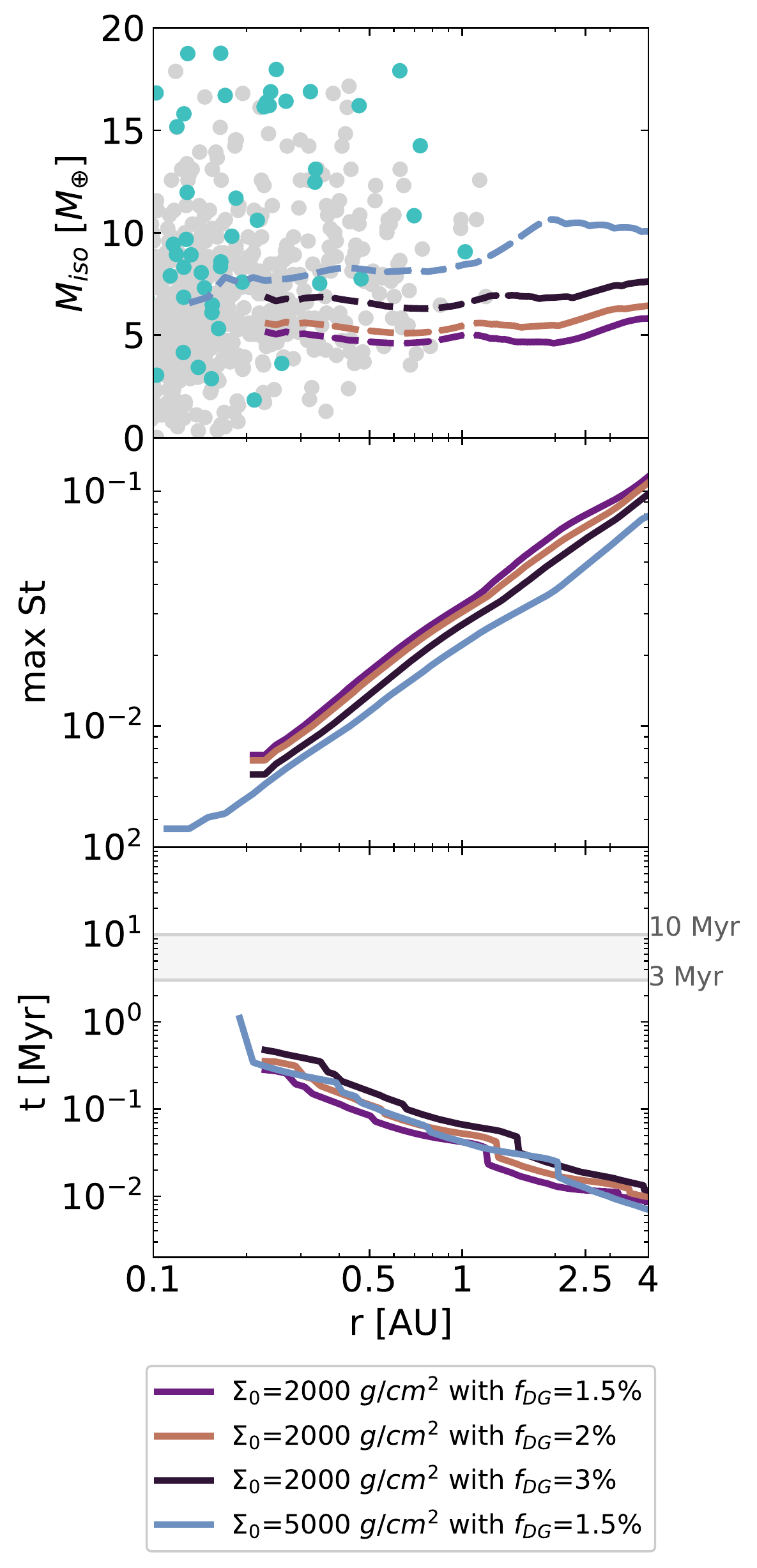}
\caption{Pebble isolation mass (top plot), maximum Stokes number (middle plot) and planetary growth timescale until the pebble isolation mass is reached (bottom plot) as a function of orbital distance, for discs with $\alpha= 10^{-3}$ and $u_f$ =  10 m/s. We show discs with higher dust-to-gas ratios, $f_{DG}=2\%$ and $f_{DG}=3\%$ and higher initial surface density $\Sigma_{g,0}$ = 5000 g/cm$^2$. We include for reference the nominal run with $\Sigma_{g,0}$ = 2000 g/cm$^2$ and $f_{DG}=1.5\%$.}
\label{Fig:maxStokes&Pebble_isolation_mass_extras}
\end{figure}

\begin{figure}
\centering
\includegraphics[width=\columnwidth]{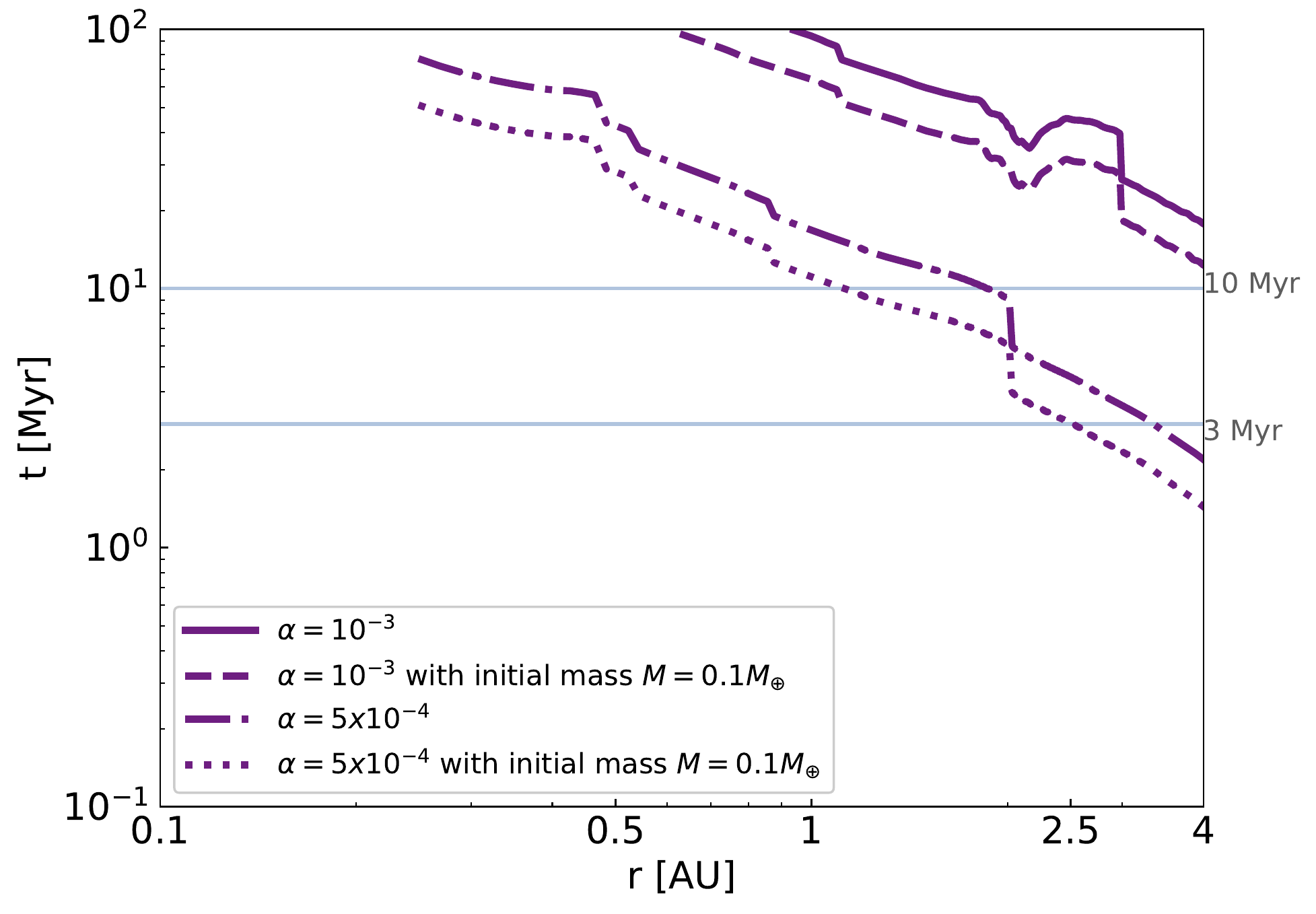}
\caption{Planetary growth timescale until the pebble isolation mass is reached as a function of orbital distance with initial planetary seed masses  $M_{init}$ = 0.1 $\rm{M_{\oplus}}$. We tested two $\alpha$-viscosities, $10^{-3}$ and $5\times10^{-4}$ with a fragmentation velocity, $u_f$ = 1 m/s. Overplotted are the nominal disc models with initial planetary seed masses  $M_{init}$ = 0.01 $\rm{M_{\oplus}}$.}
\label{Fig:timeforMiso_highinitialmass}
\end{figure}

We have concluded that the best parameters for Super-Earth formation via pebble accretion are $\alpha = 10^{-3}$ and $u_f=10$ m/s. In this case, the pebble isolation mass is reached before the dispersal of the gaseous disc regardless of the surface density and the location within the disc. The isolation mass itself depends on the surface density and with the highest surface density ($\Sigma_{g,0} = 2000$ g/cm$^2$) planets of around 5 $\rm{M_{\oplus}}$ could form by pure pebble accretion.

We investigate here how higher gas surface densities and higher dust-to-gas ratios influence the pebble isolation masses and the accretion times by pebble accretion.  We show in Fig. \ref{Fig:maxStokes&Pebble_isolation_mass_extras} (top and middle panel) the pebble isolation masses and the maximum Stokes numbers from the additional simulations. We used two different total dust-to-gas ratios, $f_{DG}=2\%$ and $f_{DG}=3\%$ with $\Sigma_{g,0} = 2000$ g/cm$^2$ and a higher gas surface density, $\Sigma_{g,0} = 5000$ g/cm$^2$ for the nominal total $f_{DG}=1.5\%$.

The difference between the nominal dust-to-gas ratio and the 2\% is very small, so the pebble isolation masses are very similar. We find slightly higher pebble isolation masses with a dust-to-gas ratio of 3\%, but the strongest improvement comes from the simulation with $\Sigma_{g,0} = 5000$ g/cm$^2$. In this case, the maximum pebble isolation masses can be around 11 $\rm{M_{\oplus}}$ near the outer boundary of the disc. 

The growth timescales (bottom panel in Fig. \ref{Fig:maxStokes&Pebble_isolation_mass_extras}) are also very similar between the simulation with the nominal total dust-to-gas ratio and the one with $f_{DG}=2\%$. If we double the amount of solids ($f_{DG}=3\%$), then the growth timescale is longer because of the larger isolation mass and the similar maximum Stokes numbers. If we increase the gas surface density to 
$\Sigma_{g,0} = 5000$ g/cm$^2$, there is enough material to accrete pebbles fast enough and reduce the growth timescale compared to the high dust-to-gas ratio simulations, however it is very similar to the nominal simulation.

In Fig. \ref{Fig:timeforMiso_highinitialmass} we show how the growth timescales change depending on the assumed initial mass of the planetary seed. We compare our nominal simulations with $\alpha = 10^{-3}$ and $\alpha = 5\times10^{-4}$ for $u_f$ = 1 m/s and an initial planetary mass of $M_{init} = 0.01$ $\rm{M_\oplus}$ with an increased initial mass of $M_{init} = 0.1$ $\rm{M_\oplus}$. We only used $\Sigma_{g,0} = 2000$ g/cm$^2$ because this surface density leads to the highest pebble isolation masses. The higher initial mass reduces the growth timescale, but the difference is small and for most of the regions of the inner discs the pebble isolation mass is not reached within the lifetime of the disc.

\section{Discussion}
\label{sec:Discussion}

\subsection{Planet growth}
\label{subsec:Planet growth}

The aspect ratios and the temperatures in our discs are relatively low, as expected by the low viscosity we have used \citep{Savvidou+2020}. The same is true for the discs with higher fragmentation velocities ($u_f $= 10 m/s in contrast to $u_f$ = 1 m/s) because the collisions are less destructive and the larger particles, which are allowed, carry lower opacities. In the context of pebble accretion, this means that the pebble isolation masses we calculate do not correlate sufficiently with the bulk of the planetary masses that are observations unless $\alpha = 10^{-3}$ and $u_f$ = 1 m/s. In this specific case, we have small enough particles, so that the opacity is sufficient to prevent significant cooling and hence high enough aspect ratios.

However, in order to have fast growth of the planets, we need either $\alpha \leq 10^{-4}$ and $u_f$ = 1 m/s or any $\alpha \leq 10^{-3}$ and $u_f$ = 10 m/s (Figs. \ref{Fig:maxStokes&timeforMiso_uf1Z0} and \ref{Fig:maxStokes&timeforMiso_uf10Z0}) since the maximum Stokes numbers are larger by up to two orders of magnitude (Figs. \ref{Fig:Pebble_isolation_mass_uf1Z0} and \ref{Fig:Pebble_isolation_mass_uf10Z0}). Therefore there seems to be a trade-off between the possible pebble isolation mass and the timescale to reach it. The most efficient parameters to have both high enough isolation masses and short enough timescales are $\alpha = 10^{-3}$ and $u_f$ = 10 m/s.

\citet{Venturini+2020a} find that $\alpha < 10^{-4}$ is needed to reach the pebble isolation mass in time. The maximum  masses they find are around 7 $\rm{M_{\oplus}}$ for an $\alpha = 10^{-5}$. They use slightly different parameters for their nominal runs (e.g. initial dust-to-gas ratio of 1\% which can increase through radial drift in the inner disc with only small difference if the initial dust-to-gas ratio is larger) and a different disc model where dust opacities do not contribute self-consistently to the disc model. Specifically, the dust opacities follow the \citet{BellLin1994} opacity law, which has been shown to produce misleading disc structures, especially when multiple grain sizes are included because they are based on micrometer sized dust \citep{Savvidou+2020}. Our model suggests that a higher viscosity is needed to prevent a too low pebble isolation mass. However, their conclusions remain consistent compared to our work, regarding that a high enough Stokes number of the particles is needed to allow fast and efficient enough growth and that more massive ice rich planets can emerge from exterior to the snow line.

We do not include gas accretion into our model, because gas accretion might not be efficient for low mass planets in the inner regions of the protoplanetary disc \citep{Ikoma+2000, Lee+2014, LambrechtsLega2017, Cimerman+2017}. Furthermore, observational constraints indicate that planets with up to 4 Earth radii have mostly only up to a few per cent of their mass in hydrogen and helium envelopes \citep{Zeng+2019}. This implies that the majority of the planetary mass for these close-in super-Earths / mini-Neptunes has to originate from solid accretion, justifying as a first approach to ignore gas accretion.

\subsection{Comparison to the Kepler data}
\label{subsec:Comparison to the Kepler data}

We find almost flat aspect ratios independently of the disc parameters used. This is an important observation as it could lead to planetary systems with similar masses. In \citet{Millholland+2017} it is suggested that similar masses in a system would also lead to similar radii, hence the constant with orbital distance aspect ratios we note in our discs support the "peas-in-a-pod" scheme \citep{Weiss+2018}. In order to reach more specific conclusions on this matter it would be important to consider multiple growing embryos at the same time.

Judging by the almost flat, slightly flaring aspect ratios, planets allowed to migrate would be expected to migrate inwards \citep{Bitsch+2015a,Savvidou+2020}. The pebble isolation masses have a low dependency on the orbital distance, hence the maximum mass planets can reach would remain unchanged within our simulations, even if we were to include planet migration. We, nevertheless, expect discs to be flared \citep[e.g.][]{ChiangGoldreich1997}, thus increasing the pebble isolation mass at larger distances to the star. The growth rate, though, is slower near the innermost regions of the discs, caused by the evaporation of solids, making growth to Super-Earth masses even harder.

It is important to include migration in future work, not only to reach more robust conclusions on the "peas-in-a-pod" configuration, but also for the role it can play in determining the composition of the planets \citep{Raymond+2018,Bitsch+2019b,Izidoro+2019}. Even though here we include several chemical species, we do not discuss the influence they could have on the planetary compositions.

Even if we only consider the pebble isolation masses that we find here (Fig. \ref{Fig:Pebble_isolation_mass_uf1Z0} and Fig. \ref{Fig:Pebble_isolation_mass_uf10Z0}), and not whether there will be time to reach them, we notice that most of the observed planetary masses are above the pebble isolation masses in our simulations. This can be explained by our self-consistent disc model with the grain size distribution leading to aspect ratios much lower than used in previous models, due to the size and composition dependent opacities. The growth of the planets, though, can be aided by collisions, either before or after the dissipation of the gas disc \citep{KominamiIda2004,OgiharaIda2009,Cossou+2014,Izidoro+2017,Ogihara+2018,Izidoro+2019}. 

The low masses that we find could also be explained if those planets are not super-Earths, but terrestrial planets. If the pebble flux is low, then we expect to have small planetary masses before the gas dissipation. Even with collisions after the gas dissipation, the masses cannot exceed a few $\rm{M_{\oplus}}$ \citep{Lambrechts+2019} which is consistent with the mass we find in this work for $\alpha = 10^{-3}$ and $u_f$ = 10 m/s. However, it is important to discuss the composition of the planets to define whether they are terrestrial or super-Earths, where planets formed during the gas phase could accrete small gaseous envelopes in contrast to planets that formed similar to the terrestrial planets via collisions after the gas phase.

\subsection{Disc parameters}
\label{subsec:Disc parameters}

We have explored a few pathways to either increase the pebble isolation masses or shorten the growth timescales. As a reminder, the fastest growth timescales with high enough isolation masses come from the models with $\alpha = 10^{-3}$ and $u_f$ = 10 m/s. The nominal dust-to-gas ratio is 1.5\%. As a consequence, we tested models with the aforementioned $\alpha$-viscosity and fragmentation velocity and the highest surface density with $\Sigma_{g,0}$ = 2000 g/cm$^2$ because this density provides the highest isolation masses. We also tested the nominal dust-to-gas ratio with higher surface density,  $\Sigma_{g,0} $ = 5000 g/cm$^2$.

We find that the higher dust-to-gas ratio, does indeed improve the isolation masses, mainly for the case with 3\%, which is twice our nominal value. However, the masses remain just above 5 $\rm{M_{\oplus}}$. The most significant improvement comes with the higher initial gas surface density. Especially near the outer boundary, the pebble isolation mass with $\Sigma_{g,0}$ = 5000 g/cm$^2$ reaches approximately 10 $\rm{M_{\oplus}}$. This implies that in order to explain the constraints of the \emph{Kepler} observations \citep{Weiss+2018} we would need very high disc masses or a significant enhancement of the dust-to-gas ratio. Local enhancements could occur in the inner regions from radial drift \citep[e.g.][]{Birnstiel+2012} or via pebble traps. These could be planet-induced pressure bumps or "traffic jams" at the evaporation fronts \citep{RosJohansen2013,IdaGuillot2016,SchoonenbergOrmel2017,DrazkowskaAlibert2017,Ros+2019}.

In addition, it is worth noting that \citet{Venturini+2020b} claim that the more massive inner super-Earths (the planets that would populate the second peak at larger radii in the radius distribution, e.g. \citet{Fulton+2017}) are actually water rich planets originating from beyond the iceline, where the pebble isolation mass is larger, in agreement with our model. Migration should be included in future work in order to determine the final positions and masses of the planets.
 
The iceline position is located around 3 AU for $\alpha = 10^{-3}$, $\Sigma_{g,0}$ = 2000 g/cm$^2$ and $u_f$ = 1 m/s. Lower $\alpha$-viscosities, gas surface densities or fragmentation velocities move the iceline inwards (see Sect. \ref{sec:Disc structure} and \citep{Savvidou+2020}). The higher dust-to-gas ratios or gas surface densities, thus, also help in keeping the iceline further out from the star. However, the position and evolution of the iceline location, along with the possibility of migration for planets defines their compositions \citep{Bitsch+2019b}. 

\section{Summary}
\label{sec:Summary}

In this work, we use the self-consistent protoplanetary disc model presented in \citet{Savvidou+2020}, with additional chemical species and the corresponding opacities (see Table \ref{tab:Grain composition} and Fig. \ref{Fig:Opacities}), focusing on the inner parts of the disc. We use the MRN grain size distribution \citep{MRN1977}, with a disc-dependent upper boundary for the grain sizes (Eq. \ref{eq:s_max}). We then combine the equilibrium disc structures from the hydrodynamical simulations with a framework to study planet growth via pebble accretion. The disc parameters we used are summarized in Table \ref{tab:Simulation_parameters}. In this work, we do not take into account planetary migration and do not discuss the planetary compositions for simplicity. Furthermore, because the growing planets are in the low mass regime we do not model gas accretion. Additionally we use only fixed disc structures in time, because our 2D model can not be evolved for several Myr.

We present the equilibrium disc structures in Figs. \ref{Fig:DiscStructure_uf1} and \ref{Fig:DiscStructure_uf10}. 
The aspect ratio profiles are almost constant with orbital distance. This is expected because at these innermost parts of the protoplanetary discs, viscous heating dominates over stellar irradiation and the disc does not flare up \citep{ChiangGoldreich1997,DullemondDominik2004,Bitsch+2015a,Ida+2016,Savvidou+2020}. This implies that the planets forming in the inner disc would have similar masses in the pebble accretion scenario, because the pebble isolation mass is a strong function of the aspect ratio, supporting the "peas-in-a-pod" scheme \citep{Millholland+2017,Weiss+2018}. 

We calculate the pebble isolation masses following the approximation by \citet{Bitsch+2018} (Figs. \ref{Fig:Pebble_isolation_mass_uf1Z0} and \ref{Fig:Pebble_isolation_mass_uf10Z0}) and then estimate the time it takes to reach them depending on the disc parameters (Figs. \ref{Fig:maxStokes&timeforMiso_uf1Z0} and \ref{Fig:maxStokes&timeforMiso_uf10Z0}). Including opacities which are grain size and composition dependent means that when the disc parameters allow large particles to form, the aspect ratios will be lower. This leads to low pebble isolation masses because they directly depend on the aspect ratio of the disc (Eq. \ref{eq:Miso}). We find the highest pebble isolation masses for $\alpha = 10^{-3}$ and $5\times 10^{-4}$ when the fragmentation velocity is $u_f$ = 1 m/s and for $\alpha = 10^{-3}$ when $u_f$ = 10 m/s, mainly for high gas surface densities, with $\Sigma_{g,0} \geq$ 1000 g/cm$^2$.

However, high pebble isolation masses also mean longer growth timescales, because of the smaller pebbles inside the disc. Comparing with the typical lifetimes of protoplanetary discs (3 to 10 Myr), we find that for low fragmentation velocities ($u_f$  = 1 m/s) the timescales for planetary growth are too long (with a small dependence on the gas surface density and the orbital distance). Hence, there is a trade-off between the pebble isolation masses and the growth timescales and we conclude that the best set of parameters is $\alpha = 10^{-3}$ and $u_f$ = 10 m/s within our model. With a gas surface density of $\Sigma_{g,0}$ = 2000 g/cm$^2$ the pebble isolation masses reach almost 6 $\rm{M_\oplus}$, which the planets can reach in less than 1 Myr.

The maximum masses that planets can reach by pure pebble accretion are relatively low and thus the masses of the majority of the observed planets can probably not be explained via pebble accretion only. We also tested higher dust-to-gas ratios and a higher surface density (Fig. \ref{Fig:maxStokes&Pebble_isolation_mass_extras}). Even though they do help in increasing the pebble isolation masses, they also bring longer or comparable timescales for planetary growth compared to the nominal simulations.

We also tested whether a higher planetary seed mass can shorten the growth timescales, by starting with $M_{init}$ = 0.1 $\rm{M_{\oplus}}$ instead of $M_{init}$ = 0.01 $\rm{M_{\oplus}}$. We find and show in Fig. \ref{Fig:timeforMiso_highinitialmass} that even though the increased initial planetary mass shortens the growth timescale, the difference is very small and growing planets still fail to reach the isolation mass within the lifetime of the disc for $\alpha = 10^{-3}$. For discs with lower alpha values, even smaller initial embryos (0.01 Earth masses) can grow fast enough to reach pebble isolation mass before the end of the disc's lifetime.

The growth of planets via pebble accretion, can be aided by collisions either before or after the dissipation of the gas \citep{KominamiIda2004,Cossou+2014,Izidoro+2017,Izidoro+2019}. It is also possible that the low pebble isolation masses we find mean that this formation mechanism leads to planet formation after gas disc dispersal rather than to planet formation during the gas disc phase. However, even with some collisions, the expected masses are not very high \citep{Lambrechts+2019}, if the initial planetary masses are small. 
However, in order to reach a definite conclusion on this, future simulations including N-body interactions are needed.

We have shown in this work that a self-consistent treatment between the pebble sizes and disc structures is of crucial importance for planet formation simulations. In particular, we find that discs which support a large pebble isolation mass, also harbor low pebble accretion rates due to the small particle sizes, hence the growth timescales can be very long.

\begin{acknowledgements}

B.B. and S.S. thank the European Research Council (ERC Starting Grant 757448-PAMDORA) for their financial support. S.S is a Fellow of the International Max Planck Research School for Astronomy and Cosmic Physics at the University of Heidelberg (IMPRS-HD).                                                                                                                                                                                                                                                                                                                                                                                                                                                                                                                                                                                                                                                                                                                                                                                                                                                                                                                                                                                                                                                                                                                                                                                                                                                                                                                                                                                                                                                                                                                                                                                                                                                                                                           

\end{acknowledgements}

\bibliographystyle{aa}
\bibliography{Savvidou+2021.bib}

\newpage
\begin{appendix}
\section{Refractive indices}
\label{app:Refractive indices}

In Table \ref{tab:Grain composition} we include the sources of the refractive indices used in order to calculate the opacities per grain size and composition as a function of temperature. The refractive indices are not only dependent on the wavelength of the incident radiation, but on the temperature of the surrounding medium, too. Hence, we find for some of the species used in this work different refractive indices for measurements at different temperatures. 

\begin{figure}
\centering
\includegraphics[width=\columnwidth]{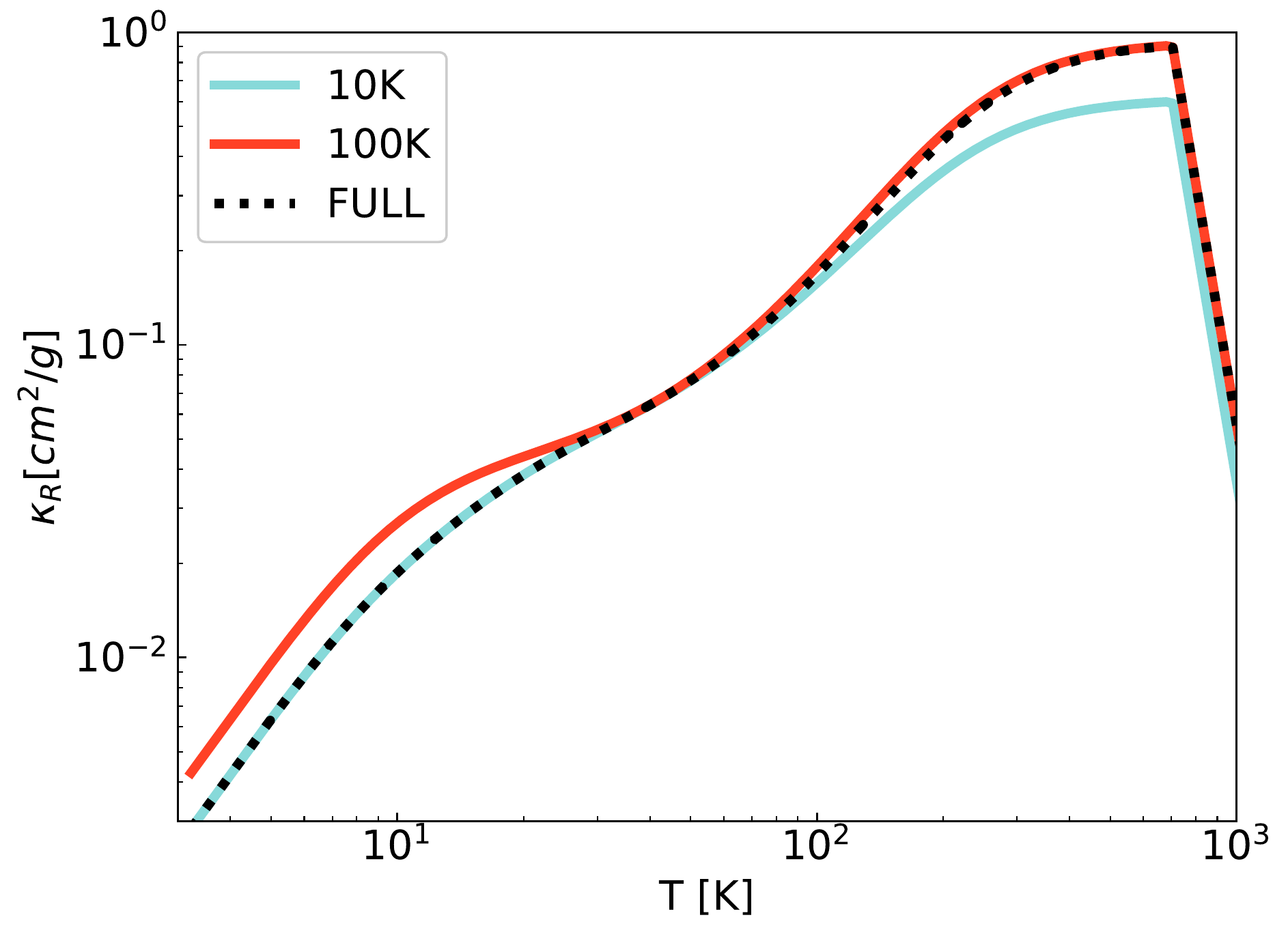}
\caption{Rosseland opacity as a function of temperature for FeS calculated using refractive indices measured at two different temperatures. The black dashed line shows the opacity we use here which combines the opacities derived from the two temperatures.}
\label{Fig:FeS}
\end{figure}

From \citet{HenningMutschke1997}, we obtain refractive indices as a function of wavelength for FeS measured at T = 10 and 100 K. In our disc models we assume that the dust temperature is the same as the temperature of the surrounding gas, which is a good approximation for the optically thick parts of the disc \citep{KampDullemond2004}. For this reason, we choose to combine the refractive indices obtained for different temperatures, so that the new refractive indices correspond to the values for T = 10 K for low temperatures but gradually switch to the values corresponding to T = 100 K for high temperatures. In Fig. \ref{Fig:FeS} we show the Rosseland opacity as a function of temperature, calculated with the refractive indices measured at T = 10 and 100 K and the combination of those which we use in the opacity module of this work.

\begin{figure*}
\centering
\begin{subfigure}{0.45\textwidth}
\includegraphics[width=\columnwidth]{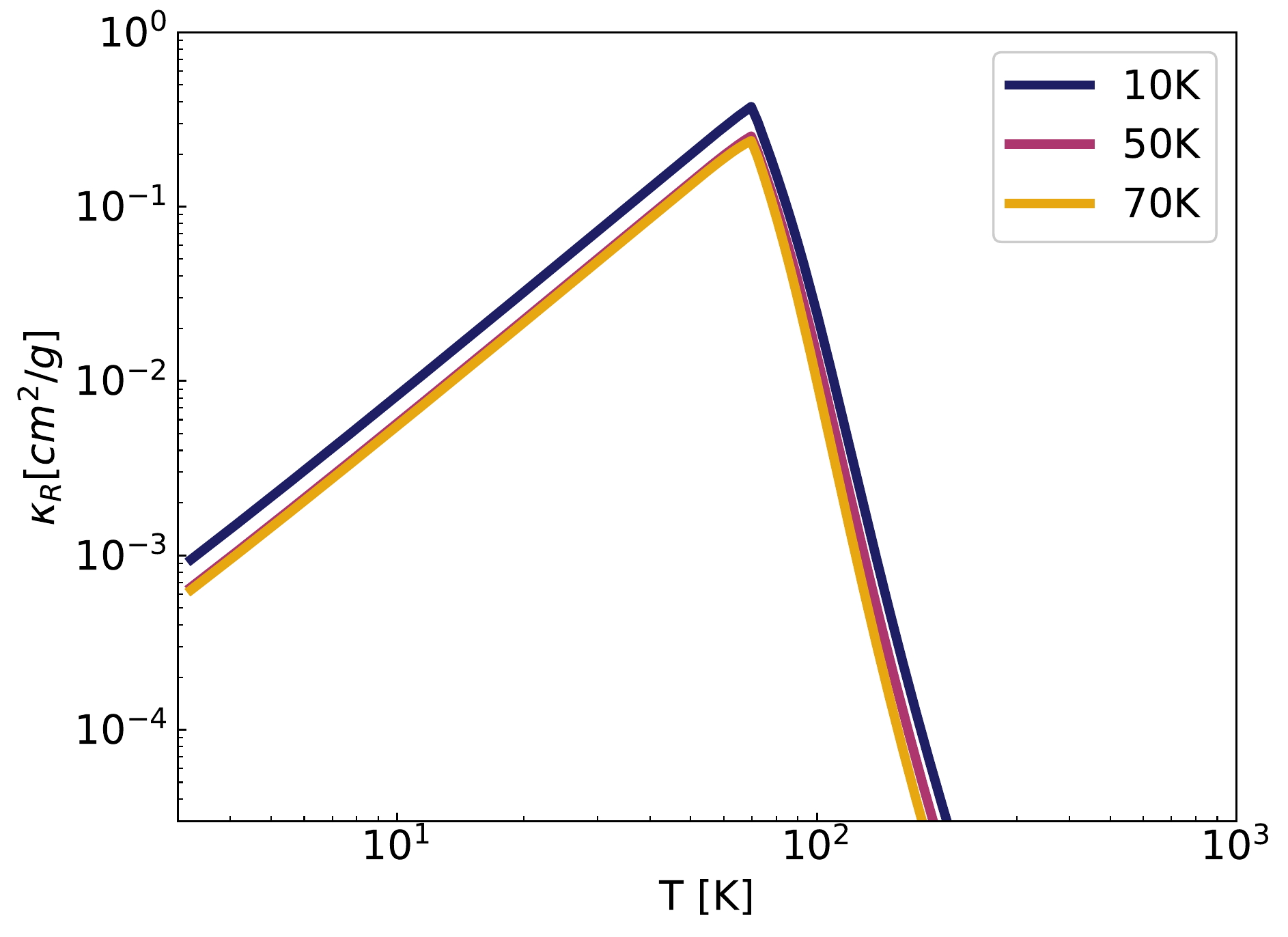}
\caption{$CO_2$}
\label{Fig:CO2}
\end{subfigure}
\begin{subfigure}{0.45\textwidth}
\includegraphics[width=\columnwidth]{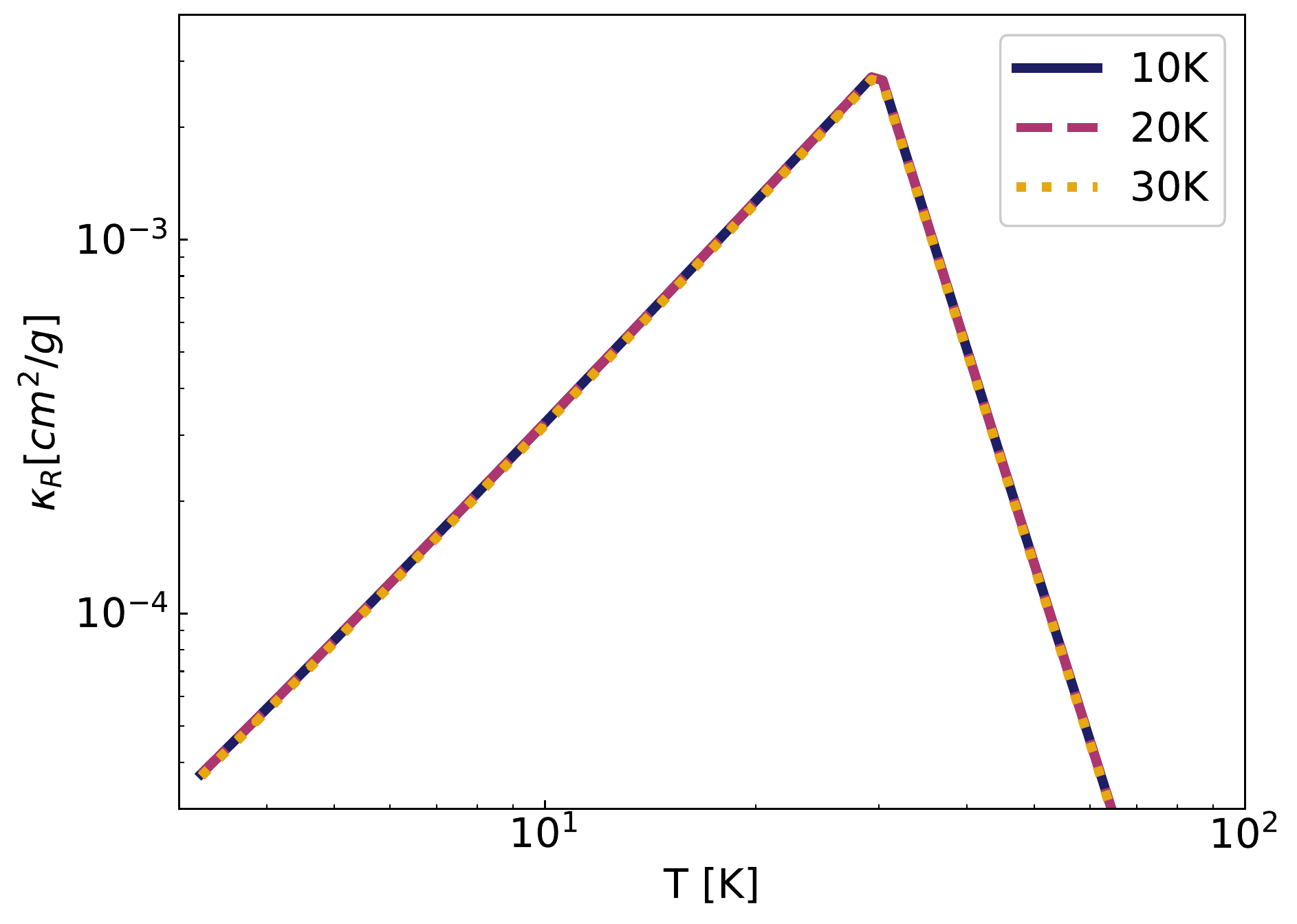}
\caption{$CH_4$}
\label{Fig:CH4}
\end{subfigure}
\caption{Rosseland opacity as a function of temperature for  the specified species calculated using refractive indices measured at three different temperatures.}
\end{figure*}

In \citet{Hudgins+1993}, the refractive indices are given for $CO_2$ for T = 10, 30, 50 and 70 K. We plot the resulting Rosseland opacities for some of those different measurements in Fig. \ref{Fig:CO2}. We choose to only use the refractive indices for 50 K in our simulations because the differences are very small between the three different sets. 

Similarly, for $CH_4$ we find refractive indices for T = 10, 20 and 30 K. The resulting mean Rosseland opacities are plotted in Fig. \ref{Fig:CH4}. They are almost the same, independently of the temperature at which the measurement was made. We chose for this work to use the refractive indices at T = 20 K. 

\end{appendix}

\end{document}